\documentclass{JHEP3}
\usepackage{amsmath}
\usepackage{amssymb}
\usepackage{epsfig}
\usepackage{latexsym}

\newcommand{\U}{\mathfrak U}
\newcommand{\F}{\mathfrak F}

\newcommand{\HH}{{\cal H}}
\newcommand{\RR}{{\mathbb R}}
\newcommand{\CC}{{\mathbb C}}
\newcommand{\NN}{{\mathbb N}}
\newcommand{\ZZ}{{\mathbb Z}}

\newcommand{\cS}{{\cal S}}

\newcommand{\ka}{{\kappa}}
\newcommand{\K}{{\rm K}}

\newcommand{\PSL} {\operatorname{PSL}}

\newcommand{\sn} {\operatorname{sn}}

\newcommand{\mat}[1] {\begin{pmatrix}#1\end{pmatrix}}
\newcommand{\bra}[1] {\left<#1\right|}
\newcommand{\ket}[1] {\left|#1\right>}
\newcommand{\braket}[2] {\left<#1\vphantom{#2}\right|
                         \left.\!\vphantom{#1}{#2}\right>}

\title{On surface states and star-subalgebras in string field theory}

\author{Ehud Fuchs\\
Max Planck Insitut f\"ur Gravitationsphysik\\
Albert Einstein Institut\\
14476 Golm, Germany\\
\email{udif@aei.mpg.de}
}
\author{Michael Kroyter\\
School of Physics and Astronomy\\
The Raymond and Beverly Sackler Faculty of Exact Sciences\\
Tel Aviv University, Ramat Aviv, 69978, Israel\\
\email{mikroyt@tau.ac.il}
}
\abstract{
We elaborate on the relations between surface states and squeezed
states. First, we investigate two different criteria for determining
whether a matter sector squeezed state is also a surface state and
show that the two criteria are equivalent.
Then, we derive similar criteria for the ghost sector.
Next, we refine the criterion for determining whether a surface state
is in $\HH_{\ka^2}$,
the subalgebra of squeezed states obeying $[S,\K_1^2]=0$.
This enables us to find all the surface states of the $\HH_{\ka^2}$
subalgebra, and show that it consists only of wedge states and
(hybrid) butterflies.
Finally, we investigate generalizations of this criterion and find
an infinite family of surface states subalgebras, whose surfaces are
described using a ``generalized Schwarz-Christoffel'' mapping.
}

\keywords{String Field Theory}
\preprint{TAUP-2777-04\\AEI-2004-068\\{\tt hep-th/0409020}}

\begin{document}

\section{Introduction}

String field theory~\cite{Witten:1986cc}
(see \cite{Taylor:2003gn} for a review of recent developments)
is a theory of an infinite set of interacting fields.
This supplies us with a huge variety of field configurations,
some of which lead to anomalous or contradicting results.
Therefore, finding the appropriate string field space, which does not
lead to contradictions, but still contains all the essential physics,
is a major task of string field
theory~\cite{Horowitz:1987yz,Erler:2003eq}.

A space with a finite number of modes is too restrictive.
Specifically, the non-trivial solutions of the string field equation
of motion
\begin{align}
\label{WittenEq}
Q_B\Psi+\Psi\star\Psi=0\,,
\end{align}
must have an infinite number of modes.
The analytical solution to the above equation is not known,
still it is obvious that it will involve an infinite number of modes,
due to the fact that states with a finite number of excited modes
do not form a subalgebra with respect to the star-product.
It is very plausible that the ``correct space of states'' is actually
a star-subalgebra, even if this assertion is somewhat less obvious
than it is for non-polynomial string field theories, such
as~\cite{Zwiebach:1992ie,Berkovits:1995ab,Okawa:2004ii}.
In fact, the star-subalgebra structure is in the heart of string field
theory. Moreover, it is clear that such a structure is vital at the
perturbative level, 
as the cubic vertex describes an interaction where from two
incoming fields $\Psi_1,\Psi_2$ we get an outgoing field
$\Psi_1\star\Psi_2$.
Understanding the structure of star-subalgebras can help us in
the search for the ``correct space of states''.

We can find star-subalgebras by looking at the form of the
three-vertex. The three-vertex can be written as a squeezed state over
a direct product of three Fock
spaces~\cite{Gross:1987ia,Gross:1987fk,Cremmer:1986if,Ohta:1986wn,Samuel:1986wp}.
Therefore, squeezed states form a subalgebra $\HH_\text{sq}$.
States in this subalgebra are represented
(for a single matter coordinate) as
\begin{equation}
\label{defMat}
\ket{\cS}=
  e^{-\frac{1}{2} a^\dagger_n S_{nm} a^\dagger_m}\ket{0}.
\end{equation}
We call $S_{nm}$ ``the defining matrix'' of the squeezed state.
Another subalgebra $\HH_\text{univ}$~\cite{Sen:1999xm,Rastelli:2000iu}
is defined by acting on the vacuum with the ghost oscillators
$b_n,c_n$ and the matter Virasoro operators $L^\text{matt}_n$.
Since the tachyon potential only involves states in this subalgebra,
the solution to the SFT e.o.m. should be in this subalgebra.
Surface states~\cite{LeClair:1989sp,LeClair:1989sj} also form a
subalgebra $\HH_\Sigma$, which is contained in the previously
mentioned subalgebras
\begin{align}
\HH_\Sigma \subset \HH_\text{sq}\,,\quad
\HH_\Sigma \subset \HH_\text{univ}\,.
\end{align}

Squeezed states are described by the infinite defining matrices.
Surface states can be described using a conformal map $f(z)$ from the
upper half plane to some Riemann surface.
Two different criteria were given for a squeezed state to be also
a surface state.
The first criterion~\cite{Fuchs:2002zz} is based on the original
relation between surface states and squeezed
states~\cite{LeClair:1989sp}. The second criterion is bases on the
integrability of the tau function~\cite{Boyarsky:2002yh}.
In section~\ref{sec:SScrit} we
prove that the two criteria are equivalent.
Then, in section~\ref{ghostCrit}, we generalize the criterion for the
ghost sector and for the twisted ghost system. We also show that
several different criteria are possible, relating different rows or
columns to the whole matrix.

Star-subalgebras can be especially useful when the star-product
in them gets a simple form.
In~\cite{Rastelli:2001hh}, the matrices $M^{rs}$, which define the
three-vertex, were diagonalized.
The basis in which these matrices are diagonal is continuous and
labeled by $-\infty<\ka<\infty$.
In this basis the squeezed state defining matrices become two
parameter functions, $S(\ka,\ka')$.
It should be noted, that while the star-product simplifies in this
basis, the form of $Q_B$, which is the kinetic operator
in~(\ref{WittenEq}) is quite
cumbersome~\cite{Douglas:2002jm,Erler:2002nr,Fuchs:2002wk,Belov:2002te}.
States with matrices diagonal in the $\ka$ basis,
\begin{align}
S(\ka,\ka')=s(\ka)\delta(\ka+\ka')\,,
\end{align}
form a subalgebra, which we call $\HH_\ka$.
This subalgebra is very limited. For example, it only allows for twist
invariant states
and the only projectors in this subalgebra are the sliver and the
identity.
The wedge states, which themselves form a subalgebra $\HH_W$,
are in this subalgebra.
In this paper we prove that they are the only surface states in this
subalgebra,
\begin{equation}
\HH_W=\HH_\ka \cap \HH_\Sigma\,.
\end{equation}

One can get a larger subalgebra by taking squeezed states
with a defining matrix block diagonal in $\pm \ka$,
\begin{align}
\label{SHK2}
S(\ka,\ka')=s_{13}(\ka)\delta(\ka-\ka')+s_2(\ka)\delta(\ka+\ka')\,.
\end{align}
This is the $\HH_{\ka^2}$
subalgebra of~\cite{Fuchs:2002zz}.
This subalgebra is already much larger than $\HH_\ka$.
A state in this subalgebra
can be represented by a function from $\ka>0$ to $\RR^3$.
This can be seen from~(\ref{SHK2})
by noting that $s_2(\ka)$ must be an even
function, while there are no conditions on $s_{13}(\ka)$.
There are also many projectors in this space.
The intersection $\HH_{\ka^2}\cap\HH_\Sigma$ contains, in addition to
the wedge states, also the butterflies~\cite{Fuchs:2002zz}.
The butterflies~\cite{Gaiotto:2001ji,Schnabl:2002ff,Gaiotto:2002kf},
unlike the wedge states, do not form a subalgebra by themselves.
The minimal subalgebra containing the butterflies is the subalgebra
$\HH_B$ of hybrid butterflies.
A hybrid butterfly is a non twist invariant state, which is the result
of star-multiplying two different (purebred) butterflies.
In section~\ref{sec:twistButt}, we introduce these states and describe
their properties and the subalgebra $\HH_B$,
as well as $\HH_{BW}$, which contains the wedges
and the butterflies.
The hybrid butterflies are (non-orthogonal) projectors and as such
they have a mid-point singularity.
In the continuous basis this translates into a singularity at $\ka=0$.
It would seem that the conformal map $f(z)$
of these states should obey $f(i)=\infty$, but this is not the case.

In section~\ref{sec:SSinHka}, our aim is to find all the
surface states in $\HH_{\ka^2}$.
In order to investigate this subalgebra 
we start with the condition for a surface state to be in
$\HH_{\ka^2}$ found in~\cite{Fuchs:2002zz}.
A variant of this condition was found in~\cite{Ihl:2003fw} for
the twisted ghost system.
Here, we show that these conditions are equivalent to a
first order differential equation.
We then study the $\PSL(2)$ properties of this equation
and find all its solutions.
The result is that the only surface states
in $\HH_{\ka^2}$ are the wedge states and the
hybrid butterflies of section~\ref{sec:twistButt}, that is,
\begin{equation}
\HH_{BW}=\HH_{\ka^2} \cap \HH_\Sigma\,.
\end{equation}

Finally, in section~\ref{sec:generalSSS} we investigate the geometric
meaning of the above differential
equation and find generalizations thereof that
are related to the Schwarz-Christoffel mapping.
The surfaces that the Schwarz-Christoffel type integrals describe
map the upper half plane
to convex polygons which may also have conical singularities.
We show that these ``generalized Schwarz-Christoffel'' maps can be
used to define an infinite number of star-subalgebras.

While this work was nearing completion the paper~\cite{Uhlmann:2004mv}
appeared, which overlaps parts of our sections~\ref{sec:TwistGhost},
\ref{sec:simpCrit} and~\ref{sec:TwistInvSol}.

\section{The equivalence of the two surface state criteria}
\label{sec:SScrit}

Before we show the equivalence of the surface state criteria for
matter sector squeezed states,
we introduce the two criteria and fix our notations.
The squeezed state defining matrix $S$ of~(\ref{defMat})
can be represented using the generating function
\begin{equation}
\label{SzwOfSnm}
S(z,w)\equiv \sum_{n,m=1}^\infty \sqrt{nm}S_{nm}z^{n-1}w^{m-1}\,.
\end{equation}
The inverse transformation is
\begin{equation}
\label{SnmOfSzw}
S_{nm}=\frac{1}{\sqrt{nm}}
  \oint \frac{dz dw}{(2\pi i)^2}\frac{S(z,w)}{z^n w^m}\,.
\end{equation}
The criterion given in~\cite{Fuchs:2002zz} is based on the fact that
for a surface state, $S(z,w)$ should have the
form~\cite{LeClair:1989sp}
\begin{equation}
\label{SzwOff}
S(z,w)=\frac{1}{(z-w)^2}-\frac{f'(-z)f'(-w)}{(f(-z)-f(-w))^2}\,,
\end{equation}
where $f(z)$ is the conformal transformation defining the surface
state in the CFT language.
Note that the first term does not contribute to the contour integral.
Its purpose is to remove the singularity at $z=w$.
In this case, $f(z)$ can be found from $S(z,w)$ using
\begin{equation}
\label{fOfSzw}
f(z)=\frac{z}{1-z\int_0^{-z}S(\tilde z,0)d\tilde z}\,.
\end{equation}
Thus, the algorithm for surface state identification is composed of
the following steps:
\begin{itemize}
\item Calculate $S(z,w)$ using~(\ref{SzwOfSnm}).
\item Find a candidate conformal transformation $f^{\mathrm{c}}(z)$
using~(\ref{fOfSzw}).
\item Substitute $f^{\mathrm{c}}(z)$ back in~(\ref{SzwOff}) and check
whether it reproduces $S(z,w)$ correctly.
\item If it does then $S$ is a surface state with the conformal map
$f(z)=f^{\mathrm{c}}(z)$, otherwise it is not a surface state.
\end{itemize}
Conformal maps, which are related by a $\PSL(2)$
transformation, describe the same state, and share the same matrix
$S$. For a given $S$, eq.~(\ref{fOfSzw}) finds in this
equivalence class the function $f(z)$ that obeys
\begin{equation}
\label{fCond}
f(0)=f''(0)=0\,,\qquad f'(0)=1\,.
\end{equation}

The conventions of~\cite{Boyarsky:2002yh} are a bit different than the
ones we use here. They write the state in their eq.~(4.22),(4.23) as
\begin{eqnarray}
\ket{N}=e^{\frac{1}{2}
  \sum_{n,m=1}^\infty\alpha_{-n} N_{n,m}\alpha_{-m}}\ket{0}\,,\\
\label{NnmOff}
N_{nm}=\frac{1}{nm}\oint_\infty
   \frac{d\tilde z d\tilde w}{(2\pi i)^2} \tilde z^n \tilde w^m
   \frac{\tilde f'(\tilde z) \tilde f'(\tilde w)}
          {(\tilde f(\tilde z)-\tilde f(\tilde w))^2}\,.
\end{eqnarray}
where as usual
\begin{equation}
\alpha_{-n}=\sqrt{n} a^\dagger_n\,,
\end{equation}
and $\tilde f(\tilde z)$ is a conformal transformation regular at
infinity.
There are three convention differences, a minus sign in the definition
of $N$, the use of the modes $\alpha_{-n}$ instead of $a^\dagger_n$
and the use of the conformal transformation $\tilde f$.
The first two differences are taken care of by setting
\begin{equation}
N_{nm}=-\frac{1}{\sqrt{nm}}S_{nm}\,.
\end{equation}

The two conformal maps can be related by a BPZ conjugation
\begin{equation}
\tilde f(\tilde z)=f(-\frac{1}{\tilde z})=f(-z)\,,
\end{equation}
where in the last equality we used $z=\tilde z^{-1}$.
Now~(\ref{SnmOfSzw}),(\ref{SzwOff}) are equivalent to~(\ref{NnmOff}).
We note that the use of a conformal transformation regular
at infinity, as well as the orientation change, are more commonly used
in the definition of a bra than of a ket. Still, the definitions
of~\cite{Boyarsky:2002yh} are consistent.
In any case,
most states considered in the literature are BPZ-real. For
these states the matrices defining the ket and bra states are the
same, and the minus sign in
the argument of $f$ in~(\ref{SzwOff}) is inessential.

The criterion found in~\cite{Boyarsky:2002yh} is based on the relation
of the matrix $N$ to the matrix of second derivatives
of the tau function of analytic curves.
The tau function obeys the Hirota identities.
The relevant identities
essentially state that the matrix of second
derivatives is determined from a single row.
These relation are summarized in eq.~(3.8) of~\cite{Boyarsky:2002yh},
which we rewrite in terms of $N$ and the $z,w$ variables
\begin{equation}
\exp\Big(\sum_{n,m=1}^\infty N_{nm} z^n w^m\Big)=
 1-\frac{\sum_{k=1}^\infty N_{1k}(z^k-w^k)}{z^{-1}-w^{-1}}\,.
\end{equation}
We can use this equation to write $N_{nm}$ in terms of its first row
as\footnote{This expression can be better understood by
writing
$x=-zw\sum_{k=1}^\infty N_{1k}(z^{k-1}+z^{k-2}w+...+w^{k-1})$
and using the Taylor expansion
$\log(1-x)=-(\sum_{l=1}^\infty\frac{x^l}{l})$.
The expression for $N_{nm}$ is a sum with coefficients
$c_{\vec k,\vec l;n,m}$
of all monomials of the form
$N_{1,{k_1}}^{l_1}\cdot...\cdot N_{1,{k_a}}^{l_a}$, where the $k_i$
are all distinct, such that ``the total index is conserved'',
and the ``total power'' is not higher than $n,m$. That is,
$l\equiv \sum_{i=1}^a l_i\leq \min(n,m)$,
$\sum_{i=1}^a l_i(k_i+1)= \, l+\vec l \cdot \vec k=n+m$.
The coefficient of this monomial is
$c_{\vec k,\vec l;n,m}=
-(-1)^l
 \frac{(l-1)!}{\prod (l_i !)}\tilde c_{\vec k,\vec l;m-l}$,
and the coefficients
$\tilde c_{\vec k,\vec l;m-l}=\frac{1}{(m-l)!}
  \partial_u^{m-l}\left.\frac{\prod_{i=1}^a
      (1-u^{k_i})^{l_i}}{(1-u)^l}\right|_{u=0}$
are solutions of the combinatorial
problem of dividing $m-l$ identical ``balls'' into $l$ boxes,
with the size of the first $l_1$ boxes adequate for at most $k_1-1$
balls, and so on. The fact that
$\tilde c_{\vec k,\vec l;m-l}=\tilde c_{\vec k,\vec l;n-l}$,
can be seen from the symmetry of interchanging ``balls'' and
``holes''.
}
\begin{equation}
\label{BRcrit}
N_{nm}=\frac{1}{(2\pi i)^2}\oint \frac{dz dw}{z^{n+1}w^{m+1}}
  \log(1-x)\,,\qquad
 x\equiv \frac{\sum_{k=1}^\infty N_{1k}(z^k-w^k)}{z^{-1}-w^{-1}}\,.
\end{equation}
It is interesting to note that
\begin{equation}
\label{Ntwist}
(N_{1n}=0 \quad n\equiv_2 0)\Rightarrow
   (N_{nm}=0 \quad n+m\equiv_2 1)\,. 
\end{equation}
Thus, a surface state is twist invariant iff
its defining matrix does not mix $a^\dagger_1$ with even creation modes.

These relations among the $N$ matrix elements may seem different from
the criterion
of~\cite{Fuchs:2002zz} described above. We now turn to prove that they
are indeed equivalent.
We rewrite~(\ref{SzwOfSnm}) as
\begin{equation}
S(z,w)\equiv -\!\!\sum_{n,m=1}^\infty nm N_{nm}z^{n-1}w^{m-1}\,.
\end{equation}
Eq.~(\ref{fOfSzw}) can be written more explicitly as
\begin{equation}
f(z)=\frac{z}{1-z h(-z)}\,,\qquad
 h(z)\equiv -\sum_{n=1}^\infty z^{n}N_{1n}\,.
\end{equation}
From here we can infer that the condition~(\ref{Ntwist}) of twist
invariance is equivalent to $f(z)$ being an odd function.
Now, eq.~(\ref{SnmOfSzw}),(\ref{SzwOff}) give $N_{nm}$ in terms of
$N_{1k}$ via
\begin{equation}
\label{finalCrit}
N_{nm}=\frac{1}{nm}\oint \frac{dz dw}{(2\pi i)^2}\frac{1}{z^n w^m}
  \frac{(1-z^2 h'(z))(1-w^2 h'(w))}{\big(z-w-zw(h(z)-h(w)\big)^2}\,.
\end{equation}
This already looks quite similar to~(\ref{BRcrit}).
Rewriting $x$ in~(\ref{BRcrit}) as
\begin{equation}
x=\frac{zw}{z-w}(h(z)-h(w))\,,
\end{equation}
and integrating by parts with respect to $z,w$, again disregarding
the term $(z-w)^{-2}$,
brings~(\ref{BRcrit}) exactly
to~(\ref{finalCrit}).
Thus, both criteria are one and the same.

\section{Surface states in the ghost sector}
\label{ghostCrit}

We now repeat the computations of the matter sector for the ghost
sector, for the regular ghost and the twisted ghost cases,
and find which ghost squeezed state matrices are related to
surface states.
The major difference is that the defining metric is not symmetric
anymore, because it relates the $b$ and $c$ ghosts.
It is also possible to consider squeezed states that are bi-linear
with respect to $b$ or $c$ oscillators. While such states can generate
symmetries in string field theory~\cite{Zwiebach:2000vc} they are not
surface states, since surface states have a definite (zero) ghost
number. Thus, in the ghost sector we consider only states of the form,
\begin{equation}
\label{ghS}
\ket{\cS}=
e^{c_{-n} S_{nm} b_{-m}}\ket{0}.
\end{equation}
Therefore, in the ghost sector, we will get two different
conditions on $S$. The difference between them is that one condition
reproduces the matrix $S$
from a single row while the other uses a single column.

Another difference of the ghost sector is that we have zero-modes that
need to be saturated. We will find a condition for both the regular
ghost sector and the twisted ghost sector. These have different
conformal weights and therefore different number of zero-modes.
The standard ghost system is a $bc$ system with conformal weights
$(h_b,h_c)$ of $(2,-1)$ and three zero-modes. The vacuum in this sector
is normalized as,
\begin{equation}
\bra{0}c_{-1} c_0 c_1\ket{0}=1\,.
\end{equation}
From here we see that the summation in~(\ref{ghS}) is in the range
$n\geq -1$, $m\geq 2$ in this case.
The twisted ghost is a $bc$ system with conformal weights
$(h_b,h_c)$ of $(1,0)$ and one zero-mode. Now, the vacuum is normalized
as,
\begin{equation}
\bra{0} c_0 \ket{0}=1\,,
\end{equation}
and the summation in~(\ref{ghS}) is in the range
$n\geq 0$, $m\geq 1$.
We treat the regular ghost system in subsection~\ref{sec:RegGhost},
and the twisted ghost system in subsection~\ref{sec:TwistGhost}.

\subsection{The regular ghost sector}
\label{sec:RegGhost}

The generating function of the defining matrix for a given surface
state in this case is~\cite{LeClair:1989sp,Maccaferri:2003rz},
\begin{align}
\label{Sghost}
S(z,w)=\frac{f'(z)^2 f'(w)^{-1}}{f(z)-f(w)}
            \left(\frac{f(w)-f(0)}{f(z)-f(0)}\right)^3
 - \frac{1}{z-w} \left(\frac{w}{z}\right)^3.
\end{align}
The generating function is related to the matrix $S$ by,
\begin{align}
S(z,w)=\sum_{n,m} (-1)^{n+m} w^{m+1} z^{n-2} S_{nm} \,,\qquad
S_{nm}=(-1)^{n+m}\oint_0 \frac{dz dw}{(2\pi i)^2}
   \frac{S(z,w)}{z^{n-1}w^{m+2}}\,.
\end{align}

We get the first condition on the ghost generating function by taking
the most dominant contribution from the expansion around $z=0$
\begin{align}
\label{Sw}
S(z,w)=\frac{1}{z^3}S_{c_1}(w) + O(\frac{1}{z^2})\,,
\qquad S_{c_1}(w)\equiv \lim_{z\rightarrow 0}z^3 S(z,w)=
  \left(w^2-\frac{f(w)^2}{f'(w)}\right)\,,
\end{align}
where we imposed~(\ref{fCond}).
We can integrate the equation above to get,
\begin{align}
\int_{w_0}^w \frac{1}{w^2-S_{c_1}(w)}dw=
  \frac{1}{f(w_0)}-\frac{1}{f(w)}\,.
\end{align}
However, we cannot set $w_0=0$, as we would like to, since the
equation is singular there. To bypass this obstacle we modify the
integrand slightly,
\begin{align}
\int_{w_0}^w \left(\frac{1}{w^2-S_{c_1}(w)}-\frac{1}{w^2}\right)dw=
  \left(\frac{1}{f(w_0)}-\frac{1}{w_0}\right)-
  \left(\frac{1}{f(w)}-\frac{1}{w}\right).
\end{align}
Now, we can take the limit $w_0\rightarrow 0$. The lower integration
limit is well defined, since $S_{c_1}(w)\approx w^4$, as can be seen
from~(\ref{Sw}). Moreover, the limit of the expression in the first
parenthesis is zero, and we get a condition on the candidate conformal
map,
\begin{align}
f^c(w)=\left(\frac{1}{w}-
  \int_0^w dw\left(\frac{S_{c_1}(w)}
    {w^4-w^2 S_{c_1}(w)}\right)\right)^{\!-1}.
\end{align}
This map obeys~(\ref{fCond}) by construction, and it
should reproduce $S(z,w)$ when plugged back into~(\ref{Sghost}),
provided $S(z,w)$ describes a surface state.

The second condition comes from expanding the generating function
around $w=0$
\begin{align}
\label{Sz}
S(z,w)=w^3 S_{b_{-2}}(z)+O(w^4)\,,\qquad
S_{b_{-2}}(z)\equiv \lim_{w\rightarrow 0}w^{-3} S(z,w)=
  \left(\frac{f'(z)^2}{f(z)^4}-\frac{1}{z^4}\right)\,.
\end{align}
We repeat the construction above and get
another expression for the candidate conformal map 
\begin{align}
f^c(z)=
\left(\frac{1}{z}-\int_0^z dz
  \left(\sqrt{S_{b_{-2}}(z)+\frac{1}{z^4}}
   -\frac{1}{z^2}\right)\right)^{\!-1}\,.
\end{align}

The first criterion that we derived gives the whole matrix in terms of
$S_{c_1}(w)$, which has the information about the $c_1$ row in the
matrix. The second condition gives the same in terms of
$S_{b_{-2}}(z)$,
which is equivalent to the $b_{-2}$ column in the same matrix.
We note that from both equations we can deduce that the
matrix element $S_{1,-2}=0$ for all surface states. That is, the
first column or row contain slightly more than enough information to
determine the matrix. We can see that it is not possible to get to a
function which obeys the initial conditions~(\ref{fCond}) when
$S_{1,-2}\neq 0$.

This curiosity is related to the fact that we chose the first row and
column. This restriction is not necessary. We could have gotten other
conditions from other rows or columns. 
In fact, this is possible also in the matter sector. This multiplicity
of conditions is not peculiar to the ghost sector. It is even
possible to get conditions from a diagonal of the matrix and other
data. However, conditions on higher rows or columns become
harder to solve and simultaneously become not restrictive enough. The
higher we go in the rows or columns the less information we can get.
This phenomenon should probably be addressed with the methods
of~\cite{Boyarsky:2002yh}.

Our goal here is a criterion for which the entire matrix is exactly
equivalent to a single row or column, provided it describes a surface
state. As the first ones are too restrictive, we examine
the criteria coming form the second row and column.
These row and column are indeed equivalent to the whole matrix.
We define $S_{c_0}(w)$ and $S_{b_{-3}}(z)$ as the next terms in the
expansions~(\ref{Sw}) and~(\ref{Sz}) respectively. In a similar way to
the above we now get,
\begin{align}
f^c(w)&=w e^{\int_0^w dw\frac{S_{c_0}(w)}{w^2-w S_{c_0}(w)}}\,,\\
f^c(z)&=\left(z^{-\frac{3}{2}}-\frac{3}{2}\int_0^z dz
  \left(\sqrt{S_{b_{-3}}(z)+z^{-5}}-z^{-\frac{5}{2}}\right)
\right)^{\!-\frac{2}{3}}.
\end{align}

\subsection{The twisted ghost system}
\label{sec:TwistGhost}

The generating function in the twisted ghost case
is~\cite{LeClair:1989sp,Gaiotto:2001ji},
\begin{align}
\label{SzwOfTwistGh}
S(z,w)=\frac{f'(z)}{f(z)-f(w)}
  \frac{f(w)-f(0)}{f(z)-f(0)}-\frac{1}{z-w}\frac{w}{z}
\,.
\end{align}
Now the relation to the matrix $S$ is given by,
\begin{align}
\label{SofTwistGh}
S(z,w)=\sum_{n,m} (-1)^{n+m} w^{m} z^{n-1} S_{nm} \,,\qquad
S_{nm}=(-1)^{n+m}\oint_0 \frac{dz dw}{(2\pi i)^2}
   \frac{S(z,w)}{z^{n}w^{m+1}}\,.
\end{align}
In the twisted ghost case the row $n=1$ and the column $m=1$ are
equivalent to the whole matrix for a surface state, as is the case
also in the matter sector.
The first row condition comes from the expansion
around $z=0$,
\begin{align}
S(z,w)=S_{c_{-1}}(w)+ O(z)\,,\qquad
 S_{c_{-1}}(w)\equiv S(0,w)=w^{-1}-f(w)^{-1}\,.
\end{align}
The condition in this case is extremely simple and involves no
integration,
\begin{align}
f^c(w)=\left(w^{-1}-S_{c_{-1}}(w)\right)^{\!-1}.
\end{align}
We do note, however, that while in the general
case~(\ref{SofTwistGh}) implies that the leading term in the expansion
is of order $O(z^{-1})$, this term is absent for surface states,
$S_{c_0}(w)=0$. That is, the coefficients of $c_0$ in the defining
matrix are zero for surface states.
This is reminiscent of the matter sector, where surface states do not
depend on the zero-mode.

The first column condition comes from the expansion around $w=0$
\begin{align}
S(z,w)=
w S_{b_{-1}}(z) + O(w^2)\,,\qquad
S_{b_{-1}}(z)\equiv \lim_{w\rightarrow 0}w^{-1}S(z,w)=
\frac{f'(z)}{f(z)^2}-\frac{1}{z^2}\,,
\end{align}
and gives the candidate conformal map
\begin{align}
f^c(z)=\left(\frac{1}{z}-\int_0^z S_{b_{-1}}(z)\right)
^{\!-1}.
\end{align}
Notice that this last condition is very similar to the
condition~(\ref{fOfSzw}) for the matter sector.

The relation between the twisted ghost and the matter sector can be
made more concrete by inserting~(\ref{SzwOfTwistGh})
into~(\ref{fOfSzw}), integrating by parts once with respect to $w$,
and comparing with~(\ref{SzwOff}). The result
is~\cite{Maccaferri:2003rz,Kling:2003sb},
\begin{align}
\label{StwGhMat}
S_{\text{matter}}=-E^{-1}S_{\text{ghost}} E\,,
\end{align}
where the matrix $E$ is defined as usual,
\begin{align}
E_{nm}=\delta_{n,m}\sqrt{n}\,.
\end{align}

\section{The hybrid butterflies}
\label{sec:twistButt}

The (regular) butterfly states are surface state rank one
projectors~\cite{Gaiotto:2001ji,Schnabl:2002ff,Gaiotto:2002kf}.
They are given by the conformal
transformation,
\begin{align}
\label{solButt}
& f=\frac{1}{\alpha}\sin(\alpha u)\,,\\
\label{buttDomain}
& 0\leq \alpha\leq 2\,,
\end{align}
where
\begin{equation}
\label{uPlane}
u\equiv \tan^{-1}(z)\,,
\end{equation}
is the $\hat z$ variable of~\cite{Gaiotto:2002kf}. We use this
parameter extensively below.

The fact that they are rank one projectors means that their wave
function factorizes to $g(l)g(r)$, where $l,r$ stand for the degrees
of freedom on the left and right half strings respectively and $g$ is
some functional of these degrees of freedom. In the half-string
formalism~\cite{Bordes:1991xh,Rastelli:2001rj,Gross:2001rk,
Furuuchi:2001df} we can represent such states by a matrix
\begin{equation}
\label{halfStrProj}
B=\ket{g}\bra{g}.
\end{equation}
Star-multiplication in the half-string basis is represented as matrix
multiplication and it is immediately seen that any state
of the form of~(\ref{halfStrProj}) is a projector up to a
normalization factor $\braket{g}{g}$. We do not care here about
normalization of the states and so we still refer to a state as a
projector if it is such up to a finite normalization, writing
\begin{equation}
P^2\simeq P\,.
\end{equation}
In fact, any state of the form
\begin{equation}
\label{TwistProj}
P=\ket{g}\bra{h}\,,\qquad \braket{h}{g}\neq 0\,,
\end{equation}
is a rank one (non-orthogonal) projector.
The result of a multiplication of a rank one projector by any state is
again a projector, provided the projection of the state in the
direction of the projector is non-zero,
\begin{equation}
\label{PSsquare}
P=\ket{g}\bra{h}\,,\quad \bra{h}S\ket{g}\neq 0\  \Longrightarrow\ 
(PS)^2\simeq (PS)\,.
\end{equation}
The resulting state has the same left component as the original
projector if the projector is on the left side of the multiplication
as in~(\ref{PSsquare}) and vice versa
if it is on the right side.
In particular, when we multiply two different orthogonal rank one
projectors, the result is a non-orthogonal projector with the left
part of the first projector and right part of the second one.
In the case of the butterflies we call the result a
``hybrid butterfly'', while referring to the original butterflies as
``purebred''.
As stated, a necessary condition for the existence of these states is
that the star-product of two distinct butterflies does not vanish. As
they are both rank one projectors it is enough to consider their inner
product. While it was shown that this overlap vanishes in the matter
sector~\cite{Fuchs:2002zz}, the total result of an overlap of any two
surface states in unity~\cite{LeClair:1989sj}.

Multiplying two hybrid butterflies gives a hybrid or a purebred
butterfly. These butterflies form a star-subalgebra $\HH_B$,
which is the minimal one containing the purebred butterflies.
It is clear that
\begin{equation}
\HH_B \subseteq \HH_{\ka^2}\cap\HH_{\Sigma}\,.
\end{equation}
We describe the hybrid butterflies from a conformal mapping point of
view in~\ref{sec:ConformalHybridButt}, and their form in the
$\ka$-basis in~\ref{sec:kappaTwistButt}.

\subsection{Conformal mapping representation of the hybrid butterflies}
\label{sec:ConformalHybridButt}

In the $u$ plane, the hybrid butterflies are represented by
the infinite strip bounded between the vertical lines in
fig.~\ref{fig:twButt}.
As the local coordinate patch should be inside this strip,
the parameters $\alpha_{l,r}$ have the usual
domain~(\ref{buttDomain}).
The conformal map describing this state is a map of this bounded
region onto the upper half plane. We use the simplest generalization
of~(\ref{solButt}),
\begin{equation}
\label{twButtStart}
f(u)=\sin(\alpha u+\delta)\,,
\end{equation}
where $\alpha$ represents some kind of a mean value of $\alpha_{l,r}$
and $\delta$ is a deviation from it.
We read the values of these constants from fig.~\ref{fig:twButt},
\begin{equation}
\label{alphaDtrans}
\alpha=\frac{2\alpha_l \alpha_r}{ \alpha_l + \alpha_r}\,,\qquad
\delta=\frac{\pi}{2}\frac{\alpha_r-\alpha_l}{\alpha_r+\alpha_l}\,.
\end{equation}
\FIGURE{
\centerline{
\epsfig{figure=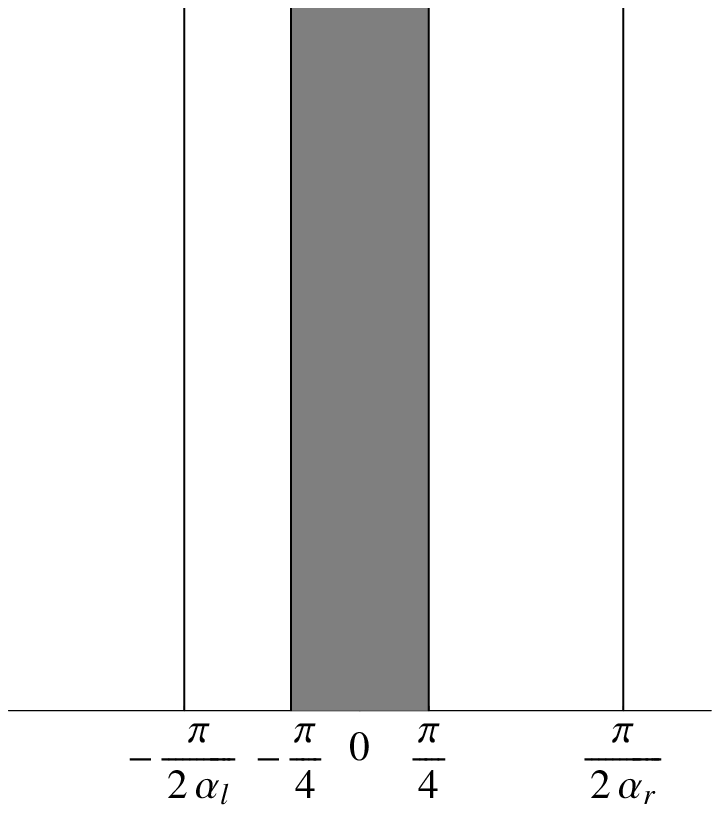,width=7.5cm}
}
\caption{A hybrid butterfly with parameters $\alpha_l$ on the left
 and $\alpha_r$ on the right, drawn in the $u$ plane.
 The local coordinate patch is in grey. It is the image of the region
 $\Im(z)\geq 0$, $|z|\leq 1$ in the $z$ plane
 under~(\ref{uPlane}).}
\label{fig:twButt}
}

While this conformal transformation describes the states well, it is
not in the standard $\PSL(2)$ form~(\ref{fCond}).
When transformed to the standard form the conformal map is
\begin{equation}
\label{twistButt}
f(u)=\frac{1}{\alpha}
 \frac{\sin(\alpha u+\delta)-\sin(\delta)}
  {\frac{\cos(2\delta)+3}{4\cos(\delta)}-
    \frac{\tan(\delta)}{2}\sin(\alpha u+\delta)}
\,.
\end{equation}
For $\delta\rightarrow 0$ this expression reduces to~(\ref{solButt}).
The restriction on the range of
$\alpha_l,\alpha_r$~(\ref{buttDomain}) implies that
\begin{equation}
\label{dAlphaRegion}
0<\alpha\leq 2\,,\qquad
0\leq \delta< \frac{\pi}{2}\,,
\end{equation}
where the cases $\alpha_l=0$ and $\alpha_r=0$, which represent
the sliver states, need to be considered separately,
as the transformation~(\ref{alphaDtrans}) is singular in this case.
In the sliver limit~(\ref{twistButt}) reduces to
\begin{equation}
\label{twistedSliver}
f(u)=\frac{u+\beta u^2}{1+\beta u+\beta^2 u^2}\,,\qquad \qquad 
  \begin{aligned}
\alpha_r&=0 &\Rightarrow \beta&=\frac{\alpha_l}{\pi} &
\quad 0<&\beta\leq 2 \\
\alpha_l&=0 &\Rightarrow \beta&=-\frac{\alpha_r}{\pi} &
\quad -2\leq&\beta<0
  \end{aligned}\,.
\end{equation}
When both $\alpha_{l,r}\rightarrow 0$ we get the sliver map,
$f(u)=u$.

We can infer the multiplication rule for the butterflies directly from
fig.~\ref{fig:twButt}. When two such states are multiplied we have to
remove the local coordinate patch from both, glue the right side of
the first with the left side of the second, and insert a local
coordinate patch between the left side of the first and the right side
of the second. We get in this way two surfaces with a single common
point at infinity. This was called a ``pinching surface''
in~\cite{Gaiotto:2002kf}, where it was explained why the half surface
without the local coordinate patch can be discarded. This
re-established the claim that
\begin{equation}
\ket{\alpha_l^1,\alpha_r^1}\star\ket{\alpha_l^2,\alpha_r^2}=
 \ket{\alpha_l^1,\alpha_r^2}\,.
\end{equation}

While these states are rank one projectors, they do not obey
$f(u=\infty)=\infty$. In~\cite{Gaiotto:2002kf} it was suggested that
rank one projectors should satisfy $f(z=i)=\infty$, which is a
manifestation of the left-right factorization of these
states. Mathematically we can understand that $f(z=i)$ is not well
defined since this is an essential singularity of~(\ref{uPlane}). For
symmetric states, as are the purebred butterflies, the directional
limit gives the correct result, while in the general case it is harder
to define it. Fig.~\ref{fig:twButt} illustrates, however, that the
left-right factorization nevertheless holds.

One more observation we can make from fig.~\ref{fig:twButt} is related
to wedge states~\cite{Rastelli:2000iu,Rastelli:2001jb,
Rastelli:2001vb,Furuuchi:2001df}.
Wedge states, like butterflies, are described by this
figure, but with the two vertical lines identified. Because of that
there is no meaning to the separate length of the left and right
parts, and there is no loss of generality by assuming that
$\alpha_l=\alpha_r$. It is common to parametrize wedge states
with the variable
\begin{equation}
\label{wedgeDomain}
n=\frac{2}{\alpha}\,,\qquad 1\leq n \leq \infty\,.
\end{equation}
With this parameter the conformal map describing wedge states is
\begin{equation}
\label{solWedge}
f(u)=\frac{n}{2}\tan(\frac{2}{n}u)\,.
\end{equation}
Wedge states form a
star-subalgebra $\HH_W$, with multiplication rule
\begin{equation}
\ket{n}\star\ket{m}=\ket{n+m-1}.
\end{equation}
In~\cite{Rastelli:2001hh} it was shown that wedge states are in
$\HH_\ka$ and so
\begin{equation}
\HH_W \subseteq \HH_\ka\cap\HH_{\Sigma}\,.
\end{equation}
We will show in section~\ref{sec:SSinHka} that this is the
whole subalgebra. Therefore, wedge states are the only surface states,
which are diagonal in the $\ka$ basis.

From the description of wedge states and butterflies using
fig.~\ref{fig:twButt}, it is immediate that the result of multiplying
a butterfly $\ket{\alpha_l,\alpha_r}$ by a wedge
state $\ket{n}$ is again a butterfly.
In this case the half surface which was previously discarded should be
glued to the right side to produce a length of
$(\alpha_r^{-1}+(n-1))\frac{\pi}{2}$ to the right of the
origin (that includes one half of the coordinate patch).
That is,
\begin{equation}
\ket{\alpha_l,\alpha_r \vphantom{(\alpha_r^{-1})^{-1}}}
\star \ket{n \vphantom{(\alpha_r^{-1})^{-1}}}=
\ket{\alpha_l,(\alpha_r^{-1}+n-1)^{-1}}.
\end{equation}
We conclude that the butterflies $\HH_B$ together with wedge states
$\HH_W$ produce again a star-subalgebra, which we label $\HH_{BW}$.
Again, it is clear that
\begin{equation}
\HH_{BW} \subseteq \HH_{\ka^2}\cap\HH_{\Sigma}\,.
\end{equation}
We will show in section~\ref{sec:SSinHka} that this is the
whole subalgebra.

\subsection{The hybrid butterflies in the $\ka$ basis}
\label{sec:kappaTwistButt}

As the hybrid butterflies are in $\HH_{\ka^2}$ we can use the methods
of~\cite{Fuchs:2002zz,Fuchs:2003wu} to find their form in the
$\ka$-basis and the half-string $\ka$-basis respectively.
We refer to these papers for technical details and conventions, and
continue briefly.
First we find $F_1(\xi)$, $F_2(\zeta)$ with the methods of section~3.3
of~\cite{Fuchs:2002zz}.
Parameterizing the states by $\alpha,\delta$ they read,
\begin{align}
\label{F1}
F_1^{\alpha,\delta}(\xi)&=
\frac{\alpha^2}{4\cosh 
  \left( \frac{\alpha \xi}{2}-i \delta\right)^2}\,,
\\
\label{F2}
F_2^{\alpha,\delta}(\zeta)&=\frac{\alpha^2}{4\sinh 
  \left( \frac{\alpha \zeta}{2}\right)^2}-\frac{1}{\sinh(\zeta)^2}\,.
\end{align}
Eq.~(\ref{F2}) does not depend on $\delta$. Thus, $s_2(\ka)$ of a
hybrid butterfly is the same as that of the appropriate purebred
butterfly,
\begin{equation}
s_2(\ka)=\cosh\left(\frac{\ka\pi}{2}\right) -
  \coth\left(\frac{\ka\pi}{\alpha}\right)
    \sinh\left(\frac{\ka\pi}{2}\right).
\end{equation}
The dependence of~(\ref{F1}) on $\delta$ is very simple and the
contour integral can be evaluated as in~\cite{Fuchs:2002zz}. Note that
no singularity can emerge from deforming the contour due to the domain
of $\delta$~(\ref{dAlphaRegion}).
The resulting matrix elements are,
\begin{equation}
\label{s13OfButt}
s_3(\ka)=s_{13}(\ka)=\frac{\sinh\left(\frac{\ka\pi}{2}\right)}
 {\sinh\left(\frac{\ka\pi}{\alpha}\right)}\,
   e^{\frac{2\ka\delta}{\alpha}}
\,,\qquad
s_1(\ka)=s_{13}(-\ka)=\frac{\sinh\left(\frac{\ka\pi}{2}\right)}
{\sinh\left(\frac{\ka\pi}{\alpha}\right)}\,
   e^{-\frac{2\ka\delta}{\alpha}}
\,.
\end{equation}
This is the representation of the hybrid butterflies in the
$\ka$-basis.

To check the special case where we have a sliver on one side we should
either substitute~(\ref{alphaDtrans}) in~(\ref{s13OfButt}) and
take the limit where either $\alpha_{l,r} \rightarrow 0$,
or find $F_{1,2}$ directly from~(\ref{twistedSliver}),
\begin{align}
\label{F1sliv}
F_1^{\beta}(\xi)&=
\frac{\beta^2}{ \left( 1+i\beta\xi\right)^2}\,,
\\
\label{F2sliv}
F_2^{\beta}(\zeta)&=\frac{1}{\zeta^2}-\frac{1}{\sinh(\zeta)^2}\,.
\end{align}
In any case we get
\begin{eqnarray}
\label{s123OfSliv}
s_2=e^{-\frac{\ka\pi}{2}}\,,&\\
s_1=\left\{ \begin{array}{cc}
 e^{-\frac{\ka}{\beta}}\sinh\left(\frac{\ka\pi}{2}\right) &
  \qquad \beta>0   \\
   0 & \qquad \beta<0
\end{array}\right . \,, \qquad \qquad &
s_3=\left\{ \begin{array}{cc}
 0 & \qquad \beta>0   \\
  e^{-\frac{\ka}{|\beta|}}\sinh\left(\frac{\ka\pi}{2}\right) &
   \qquad \beta<0
\end{array}\right . \,.\qquad\phantom{.}
\end{eqnarray}
Note that $s_{13}$ is not analytic in this limit.

To get the form of these states in the continuous half-string (sliver)
basis we use~(A.24, 3.12) of~\cite{Fuchs:2003wu} to get,
\begin{equation}
S_{\alpha_{l,r}}^h=\mat{e^{-\frac{\ka\pi}{2}a_l} & 0 \\
                         0 & e^{-\frac{\ka\pi}{2}a_r}},
\end{equation}
where we used the parametrization
\begin{equation}
a_{l,r}=\frac{2}{\alpha_{l,r}}-1\,,
\end{equation}
as in~(4.13) of~\cite{Fuchs:2003wu}.
We see that the hybrid butterflies, which are rank one projectors,
factor to left and right parts, as they should, according
to~(\ref{TwistProj}). 
We recognize these factors as the left part of the butterfly $a_l$
and the right part of the butterfly $a_r$. In fact this
could have been our starting point instead of~(\ref{twistButt}).

\section{Surface states in $\HH_{\ka^2}$}
\label{sec:SSinHka}

While searching for a squeezed state
projector, a simplifying ansatz was made
in~\cite{Kostelecky:2000hz},
\begin{equation}
[CS,\K_1]=0\,.
\end{equation}
Here $S$ is the defining matrix, $C$ is the twist matrix and $\K_1$ is
defined by the Virasoro generator $K_1=L_1+L_{-1}$ and has a
continuous spectrum $-\infty<\ka<\infty$.
These states form the $\HH_\ka$ subalgebra.
In~\cite{Fuchs:2002zz}, this ansatz was generalized by searching for
squeezed states obeying
\begin{equation}
[S,\K_1^2]=0\,.
\end{equation}
Squeezed states that are block diagonal in $\pm \ka$ form a subalgebra
of the star-product, which was dubbed $\HH_{\ka^2}$. It was found that
this space is very large and in fact equivalent to the space of
functions from $\ka>0$ to ${\RR^3}$.
It turned out that the projectors of $\HH_{\ka^2}$ consist of the
identity string field and a family of rank one projectors.
These rank one projectors are described by functions
from $\ka>0$ to ${\RR^2}$. A criterion was given for a general squeezed
state to be in $\HH_{\ka^2}$. This criterion was combined with the
criterion for a state to be a surface state~(\ref{SzwOff}) in order to
identify which surface states are in~$\HH_{\ka^2}$.
States that satisfy this condition form a subalgebra of the
star-product, as it is the intersection of two
subalgebras,~$\HH_{\ka^2}$ and $\HH_\Sigma$.
The resulting expression, although very simple and symmetric in form,
did not allow for a full description of this star-subalgebra.

This expression was later refined in~\cite{Ihl:2003fw} for the twisted
ghosts case. This should agree with the bosonic case due to the simple
relation between the three-vertices in these two
cases~(\ref{StwGhMat}).
This is so, since the space
$\HH_{\ka^2}^{\text{ghost}}$, which is the twisted ghost analogue of
$\HH_{\ka^2}$, is defined using the transformation~\cite{Erler:2002nr},
\begin{equation}
\label{contbc}
b_\ka=\sum_{n=1}^\infty \frac {v_n^\ka}{\sqrt{n}}b_n \,, \qquad
c_\ka=\sum_{n=1}^\infty v_n^\ka\sqrt{n}c_n\,,
\end{equation}
where $v_n^\ka$ are the normalized transformation coefficients from
the discrete to the continuous
basis~\cite{Rastelli:2001hh}. As this transformation is the same as
the matter one, exactly up to factors of $E,E^{-1}$, we can
use~(\ref{StwGhMat}) to deduce
that a surface state defined by a given conformal map has its matter
part in $\HH_{\ka^2}$ iff its twisted ghost part is in
$\HH_{\ka^2}^{\text{ghost}}$.
Moreover, with the proper definitions, $s_2(\ka), s_{13}(\ka)$ are the
same in both cases.

Here we further investigate surface states in $\HH_{\ka^2}$.
We simplify the criterion and solve it completely.
We then prove that
\begin{align}
\HH_\ka \cap \HH_\Sigma= &\HH_W \,,\\
\HH_{\ka^2} \cap \HH_\Sigma= &\HH_{BW}\,.
\end{align}
We start this section by simplifying the criterion
of~\cite{Fuchs:2002zz} in~\ref{sec:simpCrit}.
This will enable us to get the full solution
of the problem. First, in~\ref{sec:TwistInvSol} we find all twist
invariant states of the subalgebra.
These are the wedge states and the twist invariant
butterflies, which do not form a subalgebra by themselves.
Then, in~\ref{sec:PSL} we detour to the issue of $\PSL(2,\RR)$
invariance of the solutions, which we will use in~\ref{sec:nonTwist}
to find all possible states and conclude the proof.

\subsection{Simplifying the surface state condition}
\label{sec:simpCrit}

First, we want to simplify the condition for a surface state to be in
$\HH_{\ka^2}$. This condition is given in~(3.40)
of~\cite{Fuchs:2002zz},
\begin{equation}
\label{initEq}
  \Diamond\Box\log\left(\frac{f(z)-f(w)}{z-w}\right)=0\,,
\end{equation}
where
\begin{align}
\Box &=\frac{d^2}{du^2}-\frac{d^2}{dv^2}\,,\qquad
\Diamond =\frac{d^2}{du\,dv}\,,\\
u &=\tan^{-1}(z)\,, \qquad\ \ \ \, v =\tan^{-1}(w)\,,
\label{uvDef}
\end{align}
and we redefined $z,w \rightarrow -z,-w$ to avoid cumbersome minus
signs.
This expression can be rewritten more explicitly as
\begin{equation}
  \Box\left((1+z^2)(1+w^2)\Big(\frac{f'(z)f'(w)}{(f(z)-f(w))^2}-
     \frac{1}{(z-w)^2}\Big) \right)=0\,.
\end{equation}
Using the explicit form of the d'Alembertian
\begin{equation}
\Box=(1+z^2)\partial_z(1+z^2)
  \partial_z-(1+w^2)\partial_w(1+w^2)\partial_w\,,
\end{equation}
the condition becomes
\begin{equation}
\Big(\partial_z(1+z^2)\partial_z(1+z^2)-
    \partial_w(1+w^2)\partial_w(1+w^2)\Big)
  \Big(\frac{f'(z)f'(w)}{(f(z)-f(w))^2}- \frac{1}{(z-w)^2}\Big)=0\,.
\end{equation}

Integrating this equation $\int_0^z d z \int_0^w d w$,
with the initial conditions~(\ref{fCond}), gives
\begin{equation}
\begin{aligned}
\label{expandThis}
&(1+z^2)\partial_z(1+z^2)\partial_z \log\Big(
  \frac{f(z)-f(w)}{z-w}\frac{z}{f(z)}\Big)+\frac{1}{z^2}-\frac{1}{f(z)^2}=\\&
(1+w^2)\partial_w(1+w^2)\partial_w \log\Big(
  \frac{f(w)-f(z)}{w-z}\frac{w}{f(w)}\Big)+\frac{1}{w^2}-\frac{1}{f(w)^2}\,.
\end{aligned}
\end{equation}
This expression is already simpler than our starting point, as it has
only second derivatives as opposed to the four initial ones. However,
we can improve it further.
To that end we expand it with respect to $w$ around $w=0$, again
taking~(\ref{fCond}) into account. The first non-vanishing term is the
coefficient of $w$. It gives a necessary condition, which is a second
order differential equation with respect to $z$. Isolating
$f''(z)$, we get the not-too-illuminating expression,
\begin{equation}
\begin{aligned}
\label{fppOff}
& f''(z)=\\ & \frac{6( 2 + 2(z + z^3) f(z)f'(z) -
   2{(1 + z^2)}^2{f'(z)}^2 + {f(z)}^2(2 + f^{(3)}(0))) +
  {f(z)}^3 f^{(4)}(0)}{6{(1 + z^2)}^2 f(z)}\,.
\end{aligned}
\end{equation}
While this is already an ordinary differential equation, it is
a complicated non-linear one, and we do not know how to solve it
directly. Instead, we now re-expand~(\ref{expandThis}),
this time substituting~(\ref{fppOff}) for $f''(w)$ and
$f''(z)$. The first non-vanishing term now is the coefficient of
$w^3$, and it gives a simple first order differential equation,
\begin{equation}
\label{SimpleEq}
\frac{d f}{d u}=
(1+z^2)f'(z)=\sqrt{1+c_2 f^2+c_3 f^3 +c_4 f^4}\,,
\end{equation}
where we defined
\begin{equation}
\label{constDef}
c_2=2+f^{(3)}(0)\,,\qquad
c_3=\frac{1}{3}f^{(4)}(0)\,,\qquad
c_4=1+\frac{4}{3}f^{(3)}(0)+\frac{f^{(5)}(0)-f^{(3)}(0)^2}{12}\,,
\end{equation}
and derivatives are with respect to $z$.
It may seem that this equation is not well defined, since as a first
order equation it should depend only on one initial condition, while
we have three initial conditions in~(\ref{fCond}) and three others in
the definition of the constants~(\ref{constDef}). However, a direct
inspection shows that the equation, together with the single initial
condition $f(0)=0$, imply the other five.

So far we only proved that this equation is a necessary condition
for~(\ref{expandThis}). Strangely enough, when we
substitute~(\ref{SimpleEq}) back into~(\ref{expandThis}), we see that
it automatically holds. Eq.~(\ref{SimpleEq}) is therefore also a
sufficient condition.
We arrived at a simple criterion for a surface state to be
in~$\HH_{\ka^2}$. The defining conformal map should be a solution of a
first order differential equation in one variable with three
free parameters $c_2,c_3,c_4$.
These parameters should be real so that the coordinate patch is well
defined.
If there were no further restrictions on possible values for the
initial conditions $f^{(3)}(0),f^{(4)}(0),f^{(5)}(0)$, then
$c_2,c_3,c_4$ could have gotten any value.
We examine restrictions on these parameters and solutions of the
equation for the simple case of twist invariant states
in~\ref{sec:TwistInvSol}. Next, we examine the way
that~(\ref{SimpleEq})
generalizes when we relax the initial conditions~(\ref{fCond})
in~\ref{sec:PSL}. This will enable us to give the general solution
in~\ref{sec:nonTwist}.

We end this subsection by comparing our results to eq.~(5.16)
of~\cite{Ihl:2003fw}, which we reproduce here,
\begin{equation}
\label{IKUeq}
  -\frac{f'(u)f'(v)}{(f(u)-f(v))^2}=\frac{1-f'(u+v)}{2f(u+v)^2}-
    \frac{1+f'(u-v)}{2f(u-v)^2}\,.
\end{equation}
This is another refined version of~(\ref{initEq}).
Here, derivatives are with respect to the $u,v$
variables~(\ref{uvDef}) and not the $z,w$ ones as in our refined
expression~(\ref{expandThis}).
The initial conditions~(\ref{fCond}) take the same form with respect
to $u$ as with respect to $z$. Higher derivatives are different.
We expand~(\ref{IKUeq}) to the second order around $v=0$.
That gives us an expression for $f^{(3)}(u)$ in terms of lower
derivatives. Inserting this condition (and its derivatives) back
into~(\ref{IKUeq}), which we expand now to the third order, gives an
expression for $f''(u)$ from which the previous one can be
derived. Repeating the procedure once more, we get from the
coefficient of $v^4$ the condition~(\ref{SimpleEq}), with
\begin{equation}
c_2=f^{(3)}(0)\,,\qquad
c_3=\frac{1}{3}f^{(4)}(0)\,,\qquad
c_4=\frac{f^{(5)}(0)-f^{(3)}(0)^2}{12}\,.
\end{equation}
The reason that these conditions seem different
from~(\ref{constDef})
is that here the derivatives are with respect to $u$, while previously
they were given with respect to $z$. In fact, these conditions
coincide.
Due to the dependence on $u\pm v$ in~(\ref{IKUeq}) this equation is
not automatically satisfied upon inserting~(\ref{SimpleEq}). Rather,
we get a functional identity, which should be satisfied by all
the solutions of~(\ref{SimpleEq}), if this equation is to be a
necessary condition in this case as well.
Using the general solution~(\ref{NonTwInvSN}), found below, and
identities of Jacobi functions, we were able to prove that this is
indeed the case for all twist invariant solutions. The proof, however,
is quite messy, and will not be reproduced here. For the general case,
the expressions we get are even more ugly, and we didn't manage to
complete the proof. Instead, we were contempt by checking the
conditions numerically for various parameter values, and
expanding~(\ref{IKUeq}) up to $v^{16}$ without the emergence of any
new restrictions.

It should be noted that the condition~(\ref{IKUeq}) was derived not
for the matter sector $\HH_{\ka^2}$, but for
$\HH_{\ka^2}^{\text{ghost}}$ of the twisted ghost system.
However, we asserted above that the conditions in this two cases are
the same. Thus, the above calculation gives the expected result and
serves as a verification of the mutual consistency of our calculations
and those of~\cite{Ihl:2003fw,Maccaferri:2003rz,Erler:2002nr}.

\subsection{Twist-invariant solutions}
\label{sec:TwistInvSol}

We now examine the case of the twist invariant solutions
of~(\ref{SimpleEq}). Note that this is not a star-subalgebra.
For twist invariant solutions $f$ is anti-symmetric and thus we should
set $c_3=0$. In this case the r.h.s of~(\ref{SimpleEq}) becomes an even
function of $f(z)$ and the solution is indeed an odd function of $z$,
i.e., a twist invariant one.
Next, we write the equation in terms of $u$ as,
\begin{equation}
\label{VerySimpleEq}
\big(\frac{df}{du}\big)^2+(-c_2 f^2-c_4 f^4)=1\,.
\end{equation}
This is the energy equation of an anharmonic oscillator, with $m=2$,
$\omega^2=-c_2$ and energy $E=1$.
The role of time is played by $u$, and the
anharmonicity is described by $c_4$. For $c_2<0$ and $c_4=0$
this is the harmonic oscillator, while for $c_2,c_4<0$ we get
the Duffing oscillator.

To gain more from the oscillator analogy we use the new variable and
function
\begin{equation}
\label{Crotate}
t\equiv i u\,,\qquad g\equiv i f\,,
\end{equation}
so as to remain with the initial condition $\dot g(0)=1$.
The energy equation now is
\begin{equation}
\label{VerySimpleEqt}
\dot g^2+(c_2 g^2-c_4 g^4)=1\,,
\end{equation}
The general solution of this equation is given by the
elliptic function
\begin{equation}
\label{TwInvSN}
g=\frac{1}{k}\sn(k  t|m)\,,\qquad m=\frac{c_2}{k^2}-1\,,\qquad
k=\sqrt{\frac{c_2}{2}+\sqrt{\big(\frac{c_2}{2}\big)^2-c_4}}\,.
\end{equation}

While the solution formally exists for almost all values of the
parameters, we should keep in mind that the resulting $f$ should be a
permissible conformal map.
We know that the region $|\Re(u)| \leq \frac{\pi}{4}$, $\Im(u)\geq 0$
describes the local
coordinate patch. As such, the solution in this region should be
injective to a domain in the upper half plane.
In particular, singularities are allowed only at the boundary of the
strip. In the $t$ coordinate this strip is described by
$|\Im(t)|\leq \frac{\pi}{4}$, $\Re(t)\geq 0$.
For $g$ to be injective on the positive real line, we need that the
``particle'' will not get to a reflection point, that is, we should
not allow for cases like $a,b,c$ of fig.~\ref{fig:AnaMech}.
In particular, we should demand $c_4 \geq 0$.
The case $d$, in which $c_4>0$ and there is no potential
barrier for the particle, should not be allowed either, as in this
case the particle reaches infinity at a finite time. The only allowed
case with $c_4>0$ is $e$, as it takes infinite time to reach the
maximum point.
For $c_4=0$ we should have $c_2 \leq 0$ to avoid case $a$. We
can allow $c_2 < 0$, case $f$, as it takes infinite time to
get to infinity for the inverted harmonic potential.
We conclude that only the two classes $e,f$ are allowed solutions.
\FIGURE{
\epsfig{figure=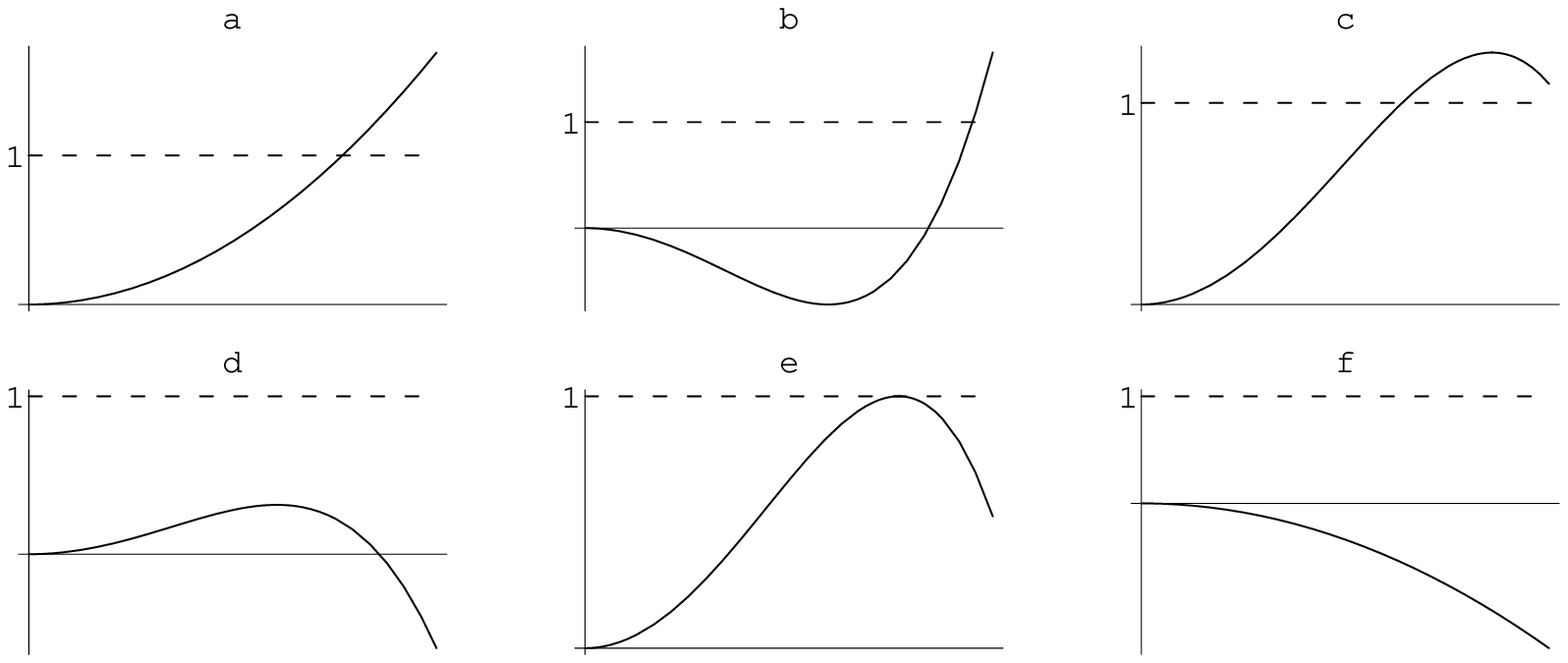,width=15cm}
\caption{Various possible ``potentials''. Cases for which infinity or
  a deflection point is reached in finite time, should be
  disregarded.}
\label{fig:AnaMech}
}

In case $e$, $c_2=2\sqrt{c_4}$, and the equation is
\begin{equation}
\dot g=1-\big(\frac{2}{n}\big)^2 g^2\,,
  \qquad n \equiv \frac{2}{\root{4}\of{c_4}}\,.
\end{equation}
The solution in terms of $f$ is the wedge state
$\ket{n}$~(\ref{solWedge}). For this state to be single valued in the
whole strip around the positive real axis,
we should further require the usual domain
restriction~(\ref{wedgeDomain}).

For case $f$ the equation is
\begin{equation}
\dot g=\sqrt{1+\alpha^2 g^2}\,,\qquad \alpha^2 \equiv -c_2\,.
\end{equation}
The solution in terms of $f$ is the butterfly~(\ref{solButt}).
Requiring injectivity in the half-infinite strip around the real
axis introduces here the usual constraint~(\ref{buttDomain}).

We see that there are no other new surface states in the twist invariant
subclass of $\HH_{\ka^2}$. In particular, the only twist
invariant surface state projectors in $\HH_{\ka^2}$ are the
butterflies and the identity (the wedge state with $n=1$), while the
only twist invariant surface states in $\HH_\ka$ are the wedge states.
In fact, all states in $\HH_\ka$ are twist invariant. Thus,
\begin{equation}
\HH_\ka \cap \HH_\Sigma=\HH_W\,.
\end{equation}

One may still wonder what goes wrong when we take a non-permissible
conformal map$f$. The answer can be seen by inspecting the butterfly
wave function in the continuous basis, eq.~(3.58)
of~\cite{Fuchs:2002zz}.
An ``illegal'' value of $\alpha$ here will result in an inverted
gaussian for the wave function.
We believe that this is the general case.
It was noticed in~\cite{Bars:2002nu,Bars:2003gu},
that the star-algebra of squeezed states
can be enlarged to give a monoid structure, so as
to allow the inclusion of inverted gaussians.
We see that surface states in this formal description do not possess a
well defined local coordinate patch.

\subsection{$\PSL(2,\RR)$ covariant differential equations}
\label{sec:PSL}

Surface states are defined as bra-states by the expectation value of
operator insertions on the given surface, which is topologically a disk.
It is customary to use the upper half plane as a canonical disk. 
This is possible due to Riemann's theorem. Surface
states are distinguished in this representation by the image
under the conformal map $f$  of the
local coordinate patch, which is the upper half unit disk.
This map sends the local coordinate patch into the upper half plane.
The boundary of the local coordinate patch is a curve in the upper
half plane, with the image of the real segment $\Im(z)=0$, $|z|<1$
lying on the boundary of the upper half
plane~\cite{Rastelli:2000iu,Rastelli:2001vb,Rastelli:2001uv,
Gaiotto:2002kf}.

In general, the state is invariant under a $\PSL(2,\mathbb{C})$ 
(M\"{o}bius) transformation
of the conformal map $f$. But since we use a canonical representation
for the disk, we are left only with a $\PSL(2,\RR)$ invariance.
Up to this point, the $\PSL(2,\RR)$ ambiguity was fixed by imposing
the conditions~(\ref{fCond}) on the map $f$.
In this subsection we want to relax
these conditions. This will help us in solving the general case
in~\ref{sec:nonTwist}, revealing an interesting mathematical structure
on the way.

We start by considering the $\PSL(2,\RR)$ transformation,
\begin{equation}
\label{sl2}
  \tilde f=\frac{a f+b}{c f+d}\,,\qquad a d-b c=1\,,
\end{equation}
which has three independent real parameters.
In terms of $\tilde f$ the l.h.s. of~(\ref{SimpleEq}) reads
\begin{equation}
\frac{d f}{du} = \frac{1}{(c\tilde f -a)^2} \frac{d \tilde f}{du}\,,
\end{equation}
while the r.h.s is
\begin{equation}
\begin{aligned}
&\sqrt{c_4 (f-f_1)(f-f_2)(f-f_3)(f-f_4)}=\\
&\sqrt{c_4 \Big(\frac{b-d\tilde f}{c\tilde f -a}-f_1\Big)
  \Big(\frac{b-d\tilde f}{c\tilde f -a}-f_2\Big)
  \Big(\frac{b-d\tilde f}{c\tilde f -a}-f_3\Big)
  \Big(\frac{b-d\tilde f}{c\tilde f -a}-f_4\Big)}\,.
\end{aligned}
\end{equation}
Here we wrote the polynomial inside the square root using its roots,
assuming the conditions
\begin{align}
\label{standCondA}
& \prod_i (-f_i) =\frac{1}{c_4}\,,\\
\label{standCondB}
& \sum_i f_i^{-1}=0\,.
\end{align}
This representation seems singular for the case of a polynomial of
degree less than four, but we will shortly see that the general case
is also covered (the minus sign in~(\ref{standCondA}) is immaterial
when the degree of the polynomial is even). In the general case $c_4$
should be replaced
by the coefficient of the highest non-vanishing power of $f$.
The reality of the coefficients of~(\ref{SimpleEq}) implies that the
roots $f_i$ are either positive, or complex conjugate paired.

We now move $(c\tilde f -a)^2$ inside the root and get,
\begin{equation}
\frac{d \tilde f}{du}=\sqrt{c_4 \prod_{i=1}^4
(c f_i+d)(\tilde f- \tilde f_i)}\,,
\end{equation}
where $f_i$ were transformed to $\tilde f_i$ using~(\ref{sl2}).
We did not use the initial conditions~(\ref{fCond})
to get to this expression.
The equation is to be supplemented with the initial condition
\begin{equation}
\tilde f(0)=\frac{a f(0)+b}{c f(0)+d}
\xrightarrow[^{^{f(0)\rightarrow0}}]{}
    \frac{b}{d}\,.
\end{equation}
We also see that when one of the roots goes to infinity, say
$\tilde f_1=\infty$, then the prefactor $c f_i+d$ exactly cancels the
divergence, and the polynomial reduces to a cubic polynomial, or to a
polynomial of a lower degree in the case of a multiple root. Given a
polynomial of degree less than four, we can transform it to a
quartic polynomial. The point which returns from
infinity contributes a factor of $(a-c\tilde f)$ to the power of its
multiplicity. The form of~(\ref{SimpleEq}) after a $\PSL(2,\RR)$
transformation is
\begin{equation}
\label{transformedSL}
\frac{d f}{d u}=\sqrt{P_4(f)}\,,\qquad
 P_4(f)=\sum_{i=0}^4 c_i f^i\,,
\end{equation}
with some initial condition $f(0)$. It is possible that some of the
coefficients vanish.

An initial condition together with a differential equation of the
form~(\ref{transformedSL}), is equivalent to a complex function, which
is a conformal transformation in our case.
These functions transform covariantly with respect to
$\PSL(2,\RR)$. The set of permissible functions of this type modulo
$\PSL(2,\RR)$ forms the $\HH_{\ka^2} \cap \HH_\Sigma$ star-subalgebra.
We want to know what are the restrictions on a real quartic
polynomial together with an initial condition in order to be
a permissible state in this subalgebra.
There is no loss of generality by assuming
that the polynomial is quartic.
We solve this problem by transforming the polynomial to its canonical
form~(\ref{standCondA}, \ref{standCondB}),
while the initial condition goes to $f(0)=0$,
by composing three simple transformations,
\begin{equation}
\begin{aligned}
f &\rightarrow f-f(0)\,, \\
f &\rightarrow \frac{f}{1-\big(\frac{1}{4}\sum_i f_i^{-1}\big)f}\,,\\
f &\rightarrow c f\,.
\end{aligned}
\end{equation}
The first of these transformations takes care of the initial condition.
The second one takes care of~(\ref{standCondB}). Note that there is no
need to find the roots of the polynomial since
\begin{equation}
\sum_i f_i^{-1}=-\frac{c_1}{c_0}\,.
\end{equation}
The third transformation is a rescaling that
fixes~(\ref{standCondA}). Since in $\PSL(2,\RR)$ there are only
positive rescalings, this transformation cannot always be preformed.
In particular, the case where $f(0)=f_i$ for some $i$, should be
excluded.
Inspecting the first two transformations, we see that the sign of the
constant term is the same as the sign of the original polynomial
evaluated at $f(0)$. Thus, we conclude that a polynomial $P(f)$, with
the initial condition $f(0)$, define a permissible function iff
\begin{equation}
\label{repCond}
P\big(f(0)\big)>0\,.
\end{equation}

\subsection{The general solution}
\label{sec:nonTwist}

Here we want to consider the case where the polynomial
in the original $\PSL(2)$ form has $c_3\neq 0$ in~(\ref{SimpleEq}).
We cannot use the simple analysis of~\ref{sec:TwistInvSol}, since in
this case we get a complex ``potential energy''.
Instead, we start by suggesting some intuition using the similarity
of~(\ref{transformedSL}) to Seiberg-Witten
curves~\cite{Seiberg:1994rs}, after which we use the $\PSL(2,\RR)$
invariance to transform the equation to a simpler form and analyze
each possible case explicitly.
We know apriori that in the twisted sector we should find the
hybrid butterflies of section~\ref{sec:twistButt}.
What we will find in the general analysis is
that they are the only permissible states in this sector. Thus, we
conclude that the subalgebra of $\HH_{\ka^2}$ surface states consists
of the wedge states and the butterflies.

In order to decide which functions are in the subalgebra, we should
impose the restriction of injectivity of the local coordinate patch.
Let us define
\begin{equation}
y=\frac{df}{du}\,,\qquad x=f\,.
\end{equation}
We can rewrite~(\ref{transformedSL}) as
\begin{equation}
y^2=P_4(x)\,,
\end{equation}
which is the elliptic equation of the Seiberg-Witten curve.
There are some differences of course. We have $\PSL(2,\RR)$ rather than
$\PSL(2,\ZZ)$ acting on our space and our surfaces have
boundaries.
However, these differences are irrelevant for the analysis below.
The topology of the Seiberg-Witten curve is that of a torus, with
\begin{equation}
\tau=\frac{\int_\alpha \frac{dx}{y}}{\int_\beta \frac{dx}{y}}\,.
\end{equation}
In our case
\begin{equation}
 \frac{dx}{y}=du\,.
\end{equation}
Therefore, going around a cut in the curve corresponds to a constant
change of $u$. As a result $f(u)$ is doubly periodic in the $u$ plane
unless a cycle collapses. But this happens exactly when two roots, or
more, coincide. Thus, it is enough to consider only these cases.

We see that we should distinguish the possible polynomials according
to the multiplicity of their roots.
The most generic case is when all roots are distinct.
We know already that this case does not contribute. We shall,
nevertheless,
analyze this case explicitly in~\ref{sec:fourRoots}, where we show
that $\PSL(2,\RR)$ can be used to get a polynomial with only even
powers, and a given initial condition.
The next case is when we have a double root.
If the two other roots are distinct, we can
transform the double root to infinity and the two other roots to
either $\pm1$ or $\pm i$, depending whether these roots are real or
not. This gives a polynomial with only constant and quadratic terms.
If there are two double roots, we can transform the two couples to
either $\pm1$ or $\pm i$, which gives again a polynomial with only
even powers.
The double root case would therefore be considered together
with the case when all roots are distinct in~\ref{sec:fourRoots},
where we verify that the distinct-roots case indeed does not
contribute, and find all the legitimate solutions to the even-power
polynomials.

Next there is the case of a triple root, which we study
in~\ref{sec:threeRoots}. In the last case, where all roots are the
same, we can send them to infinity while leaving the initial condition
invariant and rescale the constant to unity. We get the equation
\begin{equation}
\frac{d f}{d u}=1\,,\qquad f(0)=0\,,
\end{equation}
whose solution
\begin{equation}
f=u=\tan^{-1}(z)\,,
\end{equation}
is the sliver.

\subsubsection{The even-power polynomials}
\label{sec:fourRoots}

Most of the degree-four polynomials can be transformed using
$\PSL(2,\RR)$ to the form
\begin{equation}
\label{1c2c4}
P_4(f)=1+c_2 f^2+c_4 f^4\,.
\end{equation}
To show this assertion in the case of four distinct roots,
we distinguish three possible cases:
\begin{enumerate}
\item Four real roots.
\item Two real and two complex conjugate roots.
\item Two pairs of complex conjugate roots.
\end{enumerate}

In the first case we can always bring the four roots to $\pm 1$,
$x_1>1$, $-x_2<-1$, while keeping the initial condition $f(0)=0$.
We now want to $\PSL(2,\RR)$ transform the polynomial,
to make it symmetric around $f=0$ at the price of giving up $f(0)=0$.
To that end, we transform it while fixing the
points at $\pm 1$ and sending the two other points
to $\pm x$ for some $x>1$. Requiring that
such a transformation leaves $f(0)$ in the range
$|f(0)|<1$ fixes the transformation completely.
We use the cross ratio to find that
\begin{align}
x&=\frac{1 + x_1 x_2 + {\sqrt{(x_1^2-1)(x_2^2-1)}}}
   {x_1 + x_2}\,,\\
f&\rightarrow \frac{f+k}{k f+1}\,,\\
k&=\frac{-1 + x_1 x_2 - \sqrt{(x_1^2 -1)(x_2^2-1)}}{x_1-x_2}\,.
\label{4ptTrans}
\end{align}
One can see that $|k|<1$ and so the transformation is regular and
obeys all the desired conditions. A rescaling of the polynomial
brings it now to the form~(\ref{1c2c4}).
Note, however, that even in this form we cannot use the simple
analysis of~\ref{sec:TwistInvSol}, because the initial condition
$f(0)\neq 0$ becomes imaginary after the
substitution~(\ref{Crotate}).

In the second case the two real roots can be moved to $\pm 1$, while
keeping $f(0)=0$. Now the transformation should again leave the
$\pm 1$ points invariant, while sending the two conjugate points to
$\pm i x$ and leaving $|f(0)|<1$. The cross ratio determines now the
value of $x$ and then another expression for $k$~(\ref{4ptTrans}).
Again, rescaling brings us to the desired form.
Similarly, for two couples of two complex conjugate roots, we use the
cross ratio to find a value of $\theta$, such that one pair is
transformed to $e^{\pm i\theta}$, while the other is transformed to
$-e^{\mp i\theta}$. Thus, all three cases are covered.

As pointed above, the case of a double root also corresponds to a
polynomial of the form~(\ref{1c2c4}). We now turn to solve the
differential equation for this case. The problem differs from the
twist invariant case only by the initial condition. Consequently,
the general solution in this case also differs by an initial condition
constant from~(\ref{TwInvSN}),
\begin{equation}
\label{NonTwInvSN}
f=\frac{1}{k}\sn\big(k (u-u_0) |m\big),\qquad
 m= -\frac{c_2}{k^2}-1\,,\qquad
  k=\sqrt{-\frac{c_2}{2}+\sqrt{\big(\frac{c_2}{2}\big)^2-c_4}}\,.
\end{equation}
However, this function is doubly periodic in the complex plane,
\begin{equation}
\omega=4K(m)\,,\qquad \omega'=2i K(1-m)\,,
\end{equation}
where $K(m)$ is the complete elliptic integral of the first kind.
As such, this function cannot be single valued in the half-infinite
strip, unless one of the cycles diverges, or if the ratio
$\frac{K(m)}{K(1-m)}$ is imaginary, which happens only in the limit
$m \rightarrow \infty$.
We should also check the limit $k \rightarrow 0$, but this gives again
the case $m \rightarrow \infty$.
The function $K(m)$ has a branch cut, which goes from $m=1$ to
infinity. The branch cut itself does not pose a problem, as different
values for $K(m),K(1-m)$ correspond to different but equivalent
choices of lattice vectors. We are interested in the point $m=1$
itself, as this is the only point where $K(m)$ diverges.
Since the periods are proportional to $K(m),K(1-m)$, the only possible
points for a permissible conformal map are $m=0,1,\infty$.

The restriction $m=0$ implies,
\begin{equation}
\frac{c_2}{2}+\sqrt{\big(\frac{c_2}{2}\big)^2-c_4}=0\,.
\end{equation}
The only solution is $c_4=0$, $c_2<0$ that corresponds to a double
root. In this case~(\ref{NonTwInvSN}) reduces to
\begin{equation}
f(u)=\frac{\sin(\sqrt{|c_2|}u+\phi)}{\sqrt{|c_2|}}\,,
\end{equation}
where $\phi$ is fixed by the initial condition.
We recognize~(\ref{twButtStart}), which means that in this case the
solutions are the hybrid butterflies.
For $m\rightarrow\infty$, we should demand $c_4=0$, $c_2>0$.
Now, $\sin \rightarrow \sinh$, the function is not single valued in the
strip, and so this case should be excluded.

When $m=1$ we get
\begin{equation}
\big(\frac{c_2}{2}\big)^2=c_4\,.
\end{equation}
This corresponds to two double roots.
The solution now is
\begin{equation}
f(u)=\frac{\tan(\frac{\sqrt{c_2}}{2}u+\phi)}{\frac{\sqrt{c_2}}{2}}\,.
\end{equation}
For $c_2<0$ the $\tan \rightarrow \tanh$ and again this case can be
excluded as it is not single valued.
When $c_2>0$ we can $\PSL(2,\RR)$ transform the solution to the case
$\phi=0$. That is, we get only the twist invariant wedge states in
this case.

\subsubsection{The case of three identical roots}
\label{sec:threeRoots}

We choose a conformal transformation that maps the triple root
to infinity and the initial condition to $f(0)=0$, while scaling the
constant to unity. Thus, we get the equation
\begin{equation}
\frac{d f}{du}=\sqrt{1+c_1 f}\,.
\end{equation}
The solution is
\begin{equation}
f=u+\frac{c_1}{4} u^2\,.
\end{equation}
We recognize a $\PSL(2,\RR)$ transformed~(\ref{twistedSliver}). Thus,
these solutions are hybrid slivers.

We exhausted all possibilities and found no new states in the
subalgebra. We can conclude that indeed,
\begin{equation}
\HH_{BW} = \HH_{\ka^2}\cap\HH_{\Sigma}\,.
\end{equation}

\section{Other surface state subalgebras}
\label{sec:generalSSS}

The fact that a surface state is defined only up to a $\PSL(2,\RR)$
transformation implies that any star-subalgebra of surface states
should form a $\PSL(2,\RR)$ invariant subspace. With this observation
we turn to find $\PSL(2,\RR)$ invariant subspaces based on
generalizations of~(\ref{transformedSL}).
In this section, we find in this way an infinite number of surface
state subalgebras and explain their geometric interpretation.

From the way~(\ref{transformedSL}) was constructed we see that
it can be naturally generalized to
\begin{equation}
\label{GenDefEq}
\frac{d f}{du}=\big(P_n(f)\big)^{\frac{2}{n}}\,,\qquad n\in \NN\,,
\end{equation}
where $P_n(f)$ stands for a real polynomial of degree $n$ at most and the
equation is supplemented by an initial condition on $f(0)$ such
that~(\ref{repCond}) holds.
The set of solutions of this equation for a given $n$ forms a
$\PSL(2,\RR)$ invariant subspace.
We want to examine whether this
space modulo $\PSL(2,\RR)$ also forms a star-subalgebra.
Indeed the answer to this question is affirmative.
As in the case of $n=4$ analyzed above,
what we should consider is not the whole invariant subspace,
but its subset of permissible conformal maps.

Before we look at the general case we want to check the simplest
examples. For $n=1$ the equation is immediately solved. There is only
one solution, which up to $\PSL(2,\RR)$ is $f=u$, that is the sliver.
This definitely forms a
subalgebra as the sliver is a projector.
The $n=2$ case is not much harder. We can see that the solutions are
just the wedge states, again a subalgebra. The $n=3$ case already
gives beta functions in the general solution. However, it would become
evident that the only permissible solution is again the sliver.
The $n=4$ case was
the subject of section~\ref{sec:SSinHka}, where we showed that it
gives the subalgebra $\HH_{BW}$.

From checking the simplest cases it seems that the invariant subspaces
indeed form subalgebras, which we shall label $\HH_n$.
We prove below that all the $\HH_n$ are subalgebras.
We also notice that $\HH_1 \subset \HH_2 \subset \HH_4$.
It can be seen from~(\ref{GenDefEq}) that this is a part of a general
scheme,
\begin{equation}
\label{n|m}
n|m \Rightarrow \HH_n \subset \HH_m\,.
\end{equation}
That is, an element of $\HH_n$ is also a member of $\HH_m$
whenever $m$ is an integer multiple of $n$.
From here we see that we can define yet another subalgebra
\begin{equation}
\HH_\infty \equiv \bigcup_{n\in \NN} \HH_n\,.
\end{equation}
This is a subalgebra because an element of $\HH_\infty$ is necessarily
an element of $\HH_n$ for some $n$. Two arbitrary elements of
$\HH_\infty$, say $\phi_1\in \HH_n$ and $\phi_2\in \HH_m$, are both
in $\HH_{nm}$ and so is their product,
\begin{equation}
(\phi_1\star\phi_2)\in \HH_{nm} \subset \HH_\infty\,.
\end{equation}

To understand the nature of these subalgebras we recall that the
function $f$ represents a mapping from a domain in the $u$ plane to
the upper half $f$ plane. The mapping does not have to
be injective,
except for the local coordinate patch, but as it represents a disk it
should be onto, and an inverse mapping should exist. The inverse
mapping does not have to be injective either, as it may happen that
$u$ is not an adequate global coordinate for the disk. In this case
the original mapping is formally multivalued and is single valued
from some cover of the $u$ plane domain to $f$. This cover should
include a copy of the local coordinate patch and map bijectively to
the $f$ plane. We also remark that generally
the map is not restricted to the upper half $u$ plane.

We consider the inverse map from $f$ to $u$. According
to~(\ref{GenDefEq}) this map obeys
\begin{equation}
\label{uoff}
u=u_0+c \int_0^f d f
\Big((f-f_1)^{k_1}...(f-f_m)^{k_m}\Big)^{\!-\frac{2}{n}}\,,
  \qquad  k\equiv \sum_{i=1}^m k_i \leq n\,.
\end{equation}
Using $\PSL(2,\RR)$ we can put the mapping into the standard
form~(\ref{fCond}). For the map from $f$ to $u$ these conditions can
be written as
\begin{equation}
\label{uCond}
u(0)=u''(0)=0\,,\qquad u'(0)=1\,,
\end{equation}
that is
\begin{equation}
u_0=0\,,\qquad 
   c= \Big(\prod_{i=1}^m (-f_i)^{k_i}\Big)^{\frac{2}{n}}\,,\qquad
      \Big(\prod_{i=1}^m f_i^{k_i}\Big)
          \Big(\sum_{j=1}^m \frac{1}{k_j f_j}\Big)=0\,,
\end{equation}
so that
\begin{equation}
\label{PnPSL}
P_n(f)=\prod_{i=1}^m \Big(1-\frac{f}{f_i}\Big)^{k_i}=
  1+c_2 f^2+...+c_k f^k\,,
\end{equation}
as in the $P_4$ case.

The roots of a real polynomial are either real or complex paired.
In~\ref{sec:SCstates} we analyze the case when all the roots are
real. We show that these states are all projectors. We also prove that
these projectors form subalgebras, which we call $\HH_n^{(0)}$, with
$\HH_1^{(0)}=\HH_2^{(0)}$ containing only the sliver and
$\HH_4^{(0)}=\HH_{B}$. Next, in~\ref{sec:genSCstates} we analyze the
general case, where we show that the $\HH_n$ are also subalgebras.
The $\HH_n$ extend the previous ones,
$\HH_n^{(0)}\subset\HH_n$ and contain surface states with conical
singularities. The proof of the subalgebra property is
non-trivial in this case and is
completed in the appendix.

\subsection{The Schwarz-Christoffel states}
\label{sec:SCstates}

In the real root case the
expression~(\ref{uoff}) can be identified as a
Schwarz-Christoffel mapping~\cite{Driscoll}.
The image of the real $f$ axis, which is the disk boundary, is
therefore a polygon. The vertices of the polygon are the images of the
roots $f_i$. We refer to a root also as a prevertex.
The turning angle at the $i^{\text{th}}$ vertex is set by the
multiplicity $k_i$ of the prevertex as
\begin{equation}
\beta_i \pi = 2\pi \frac{k_i}{n}\,,
\end{equation}
and the interior angle $\alpha_i \pi$ between two lines which meet at
the vertex is given by
\begin{equation}
\alpha_i=1-\beta_i=1-2 \frac{k_i}{n}\,.
\end{equation}
We see that when $k=n$ the total turning angle of the polygon is
$2\pi$.
Otherwise the missing angle is at infinity.
We also see that $\beta_i>0$. Thus, the polygon is convex.

When the multiplicity of a root obeys
\begin{equation}
\label{heavyRoot}
k_i \geq \frac{n}{2}
\end{equation}
the image of the root is $u(f_i)=\infty$, whether $f_i$ is finite on
infinite, as can be seen from~(\ref{uoff}).
We have to allow for at least one such
prevertex, as otherwise the $f$ plane would be mapped to a bounded
polygon, and in particular its image would not include the local
coordinate patch.
Suppose now that there are two such prevertices. This is possible only for
even $n$ and then we get the $\tanh(c u)$ solution,
which was discarded in section~\ref{sec:SSinHka},
because it is not single valued.
Therefore, there must be a single prevertex of this type.

We conclude that the general form of a state in $\HH_n^{(0)}$ is
similar to the example shown in fig.~\ref{fig:H12}.
Conversely, given a region bounded by a polygon in the upper
half $u$ plane, which includes the local coordinate patch and with all
turning angles equal to an integer multiple of $\frac{2\pi}{n}$ for
some $n$, there is a state in $\HH_n^{(0)}$ which corresponds to this
region.
This is so because it is always possible to find a Schwarz-Christoffel
mapping to this region, with adequate powers for the
prevertices~\cite{Driscoll}.
Note that for these states a subspace of the upper $u$ plane
serves as a global coordinate. The local coordinate patch separates
the left and right sides of these states, which are therefore rank one
projectors, as stated.
\FIGURE{
\epsfig{figure=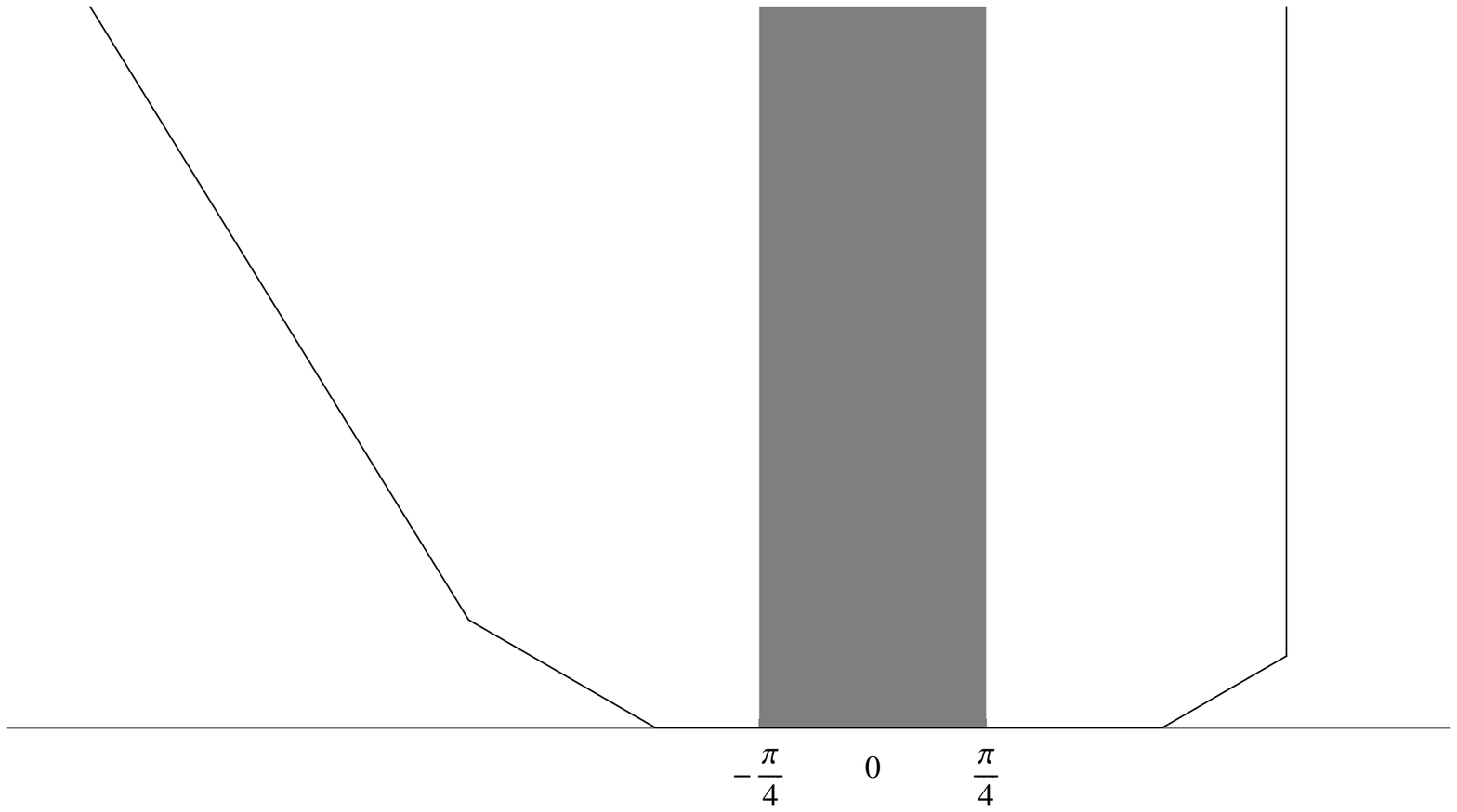,width=15cm}
\caption{The $u$ plane representation of a generic state in
  $\HH_{12}^{(0)}$. The local coordinate patch is in grey. All turning
  angles $\beta_i \pi$ are integer
  multiples of $\frac{\pi}{6}$. At infinity
  $\beta_1=\frac{7}{6}$. From there in anticlockwise direction
  $\beta_{2,3,4}=\frac{1}{6}$ and $\beta_5=\frac{1}{3}$.}
\label{fig:H12}
}

\subsection{Generalized Schwarz-Christoffel states}
\label{sec:genSCstates}

Suppose now that there exist at least a single, presumably multiple,
complex pair. The simplest case, in which the roots have maximal
possible multiplicity $\frac{n}{2}$, is possible only for an even $n$.
In this case~(\ref{uoff}) reduces to
\begin{equation}
u=c \int df(f-f_1)^{-1}(f-\bar f_1)^{-1}\,,
\end{equation}
which gives the wedge states as solutions.
We see that the root $f_1$ introduces a logarithmic
singularity in the $f$ plane. A cut should be introduced in the $f$
plane and both sides of this cut transform to different lines in the
$u$ plane. In order to remain with a disk topology these two lines in
the $u$ plane should be identified and, as a result, the point
\begin{equation}
\label{wedgef_1}
u(f_1)=\infty\,,
\end{equation}
which is the string mid-point, develops a conical singularity. This
conical singularity
carries also to the $z$ plane, except for the vacuum state $n=2$.
That the conical singularity differs in the $z,u$ coordinates can be
traced again to the essential singularity of the map~(\ref{uPlane})
for $z=i$.

This conical singularity is best described in the $\hat w$ plane
of~\cite{Rastelli:2001vb}, where it is seen that the state $\ket{n}$
has an excess angle of $(n-2)\pi$, in a representation where the local
coordinate patch takes a canonical (up to $\PSL(2,\CC)$) form.
Due to~(\ref{wedgef_1}) the two identified lines are in different
sides of the local coordinate patch and there is no left-right
separation due to the identification. Indeed, the wedge states are
generally not projectors.

In the general case, when the multiplicity of a complex root is less
than $\frac{n}{2}$, the image of the root is a finite point in
the $u$ plane. To have a non-bounded polygon in this case, we need to
demand, as in the real-root case, that there exists a prevertex
obeying~(\ref{heavyRoot}).
Also, in this case the identified curves are located on the
same side of the local coordinate patch. There is a left-right
factorization for these states and so they are projectors. Thus, the
only non-projectors in $\HH_\infty$ are the wedge states.

The addition of complex conjugate root pairs does not change the
validity of the Schwarz-Christoffel construction since on the real $f$
axis such a pair contributes a factor of
\begin{equation}
\Delta \arg\big(u'(f)\big)=
  \arg\Big(\big((f-f_i)(f-\bar f_i)\big)^{-\frac{2k_i}{n}}\Big)=0\,.
\end{equation}
This equality stems from the positivity of the factor inside the
parentheses.

Note that the introduction of real prevertices, which we did
not have for the wedge states,
may result in a curved image of a straight cut.
However, the angle between the images of the two sides of the cut
is constant along the cut and equals
\begin{equation}
\gamma_i \pi =4\pi \frac{k_i}{n}\,.
\end{equation}
The angle at infinity is reduced by this amount.
It should be possible to pick up a curved line in the $f$ plane as
the cut, in such a way that the identified lines in the $u$ plane are
straight and meet at the angle $\gamma_i \pi$.
As these line segments 
should not touch neither the local coordinate
patch nor the disk boundary, which is bounded by the real axis from
below, $\gamma>\frac{1}{2}$ is not allowed, and so states with complex
roots exist only in $\HH_{n\geq 8}$. For $\HH_{n<8}$ we have
\begin{equation}
\HH_n=\left\{ \begin{array}{ll}
  \HH_n^{(0)} & \qquad n\ \text{odd}\\ 
    \HH_n^{(0)}\cup \HH_W & \qquad n\ \text{even}
  \end{array}\right . . 
\end{equation}
It is possible to have more than one cut in the same side of the local
coordinate patch, with the restriction that the total angle of the
cuts together with the total turning angle at that side does not
exceed $\frac{\pi}{2}$.
Some examples of states are illustrated in fig.~\ref{fig:H8_2Cuts} and
in fig.~\ref{fig:H39}.\suppressfloats[t]
\FIGURE[t]{
\epsfig{figure=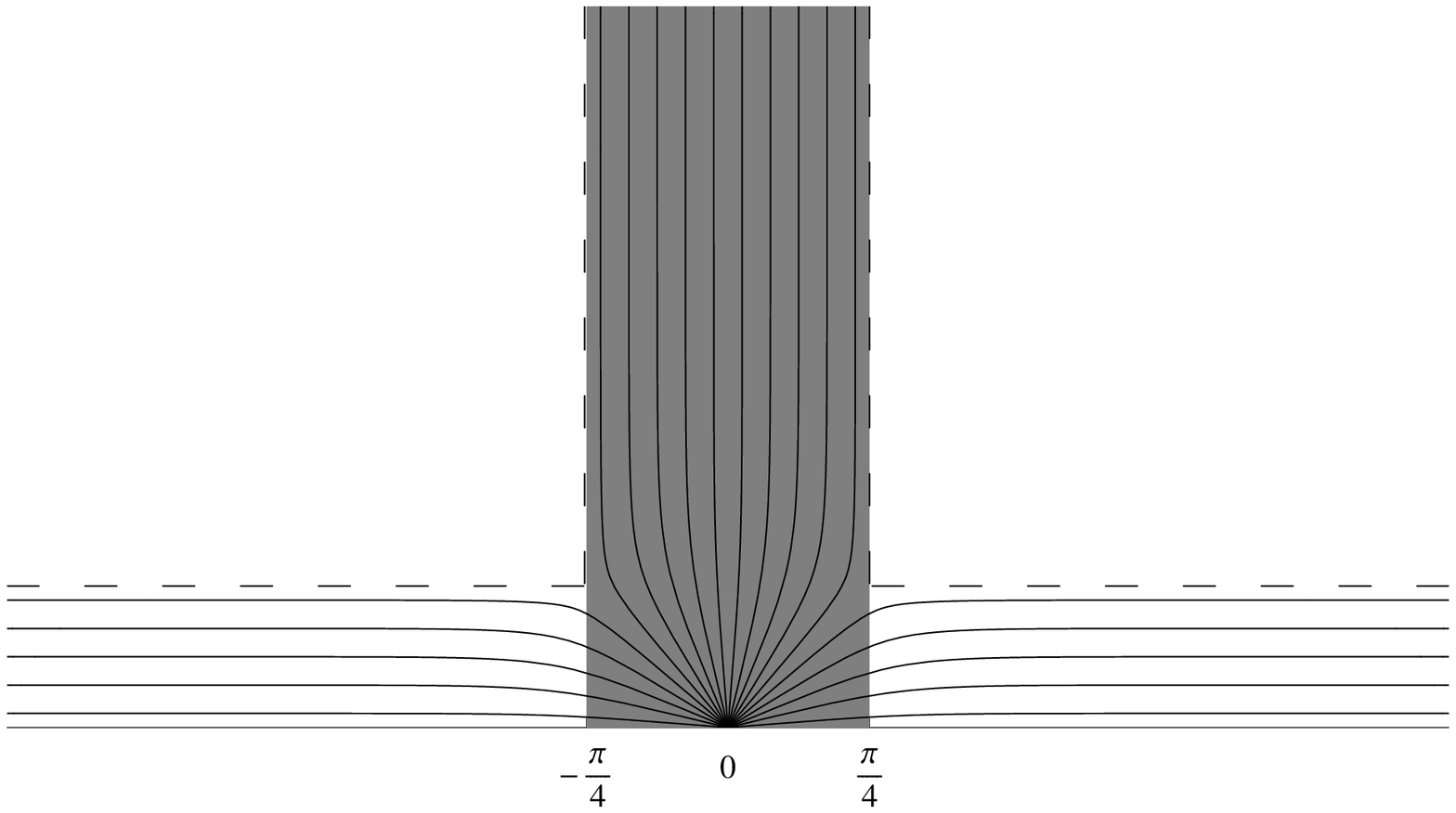,width=12.5cm}
\caption{The $u$ plane representation of the state formed in
  $\HH_{8}$ by two complex paired roots at
  $\frac{\pm 1\pm i}{\sqrt{2}}$ (and four roots at infinity), 
  that is $P_8(f)=1+f^4$. The local coordinate patch is in grey. The
  lines represent the images of radial lines, separated in the $f$
  plane by an angle of $\frac{\pi}{20}$ from each other.
  The horizontal dashed line on the left is identified with the left
  boundary of the local coordinate patch, and similarly
  on the right.}
\label{fig:H8_2Cuts}
}
\FIGURE{
\epsfig{figure=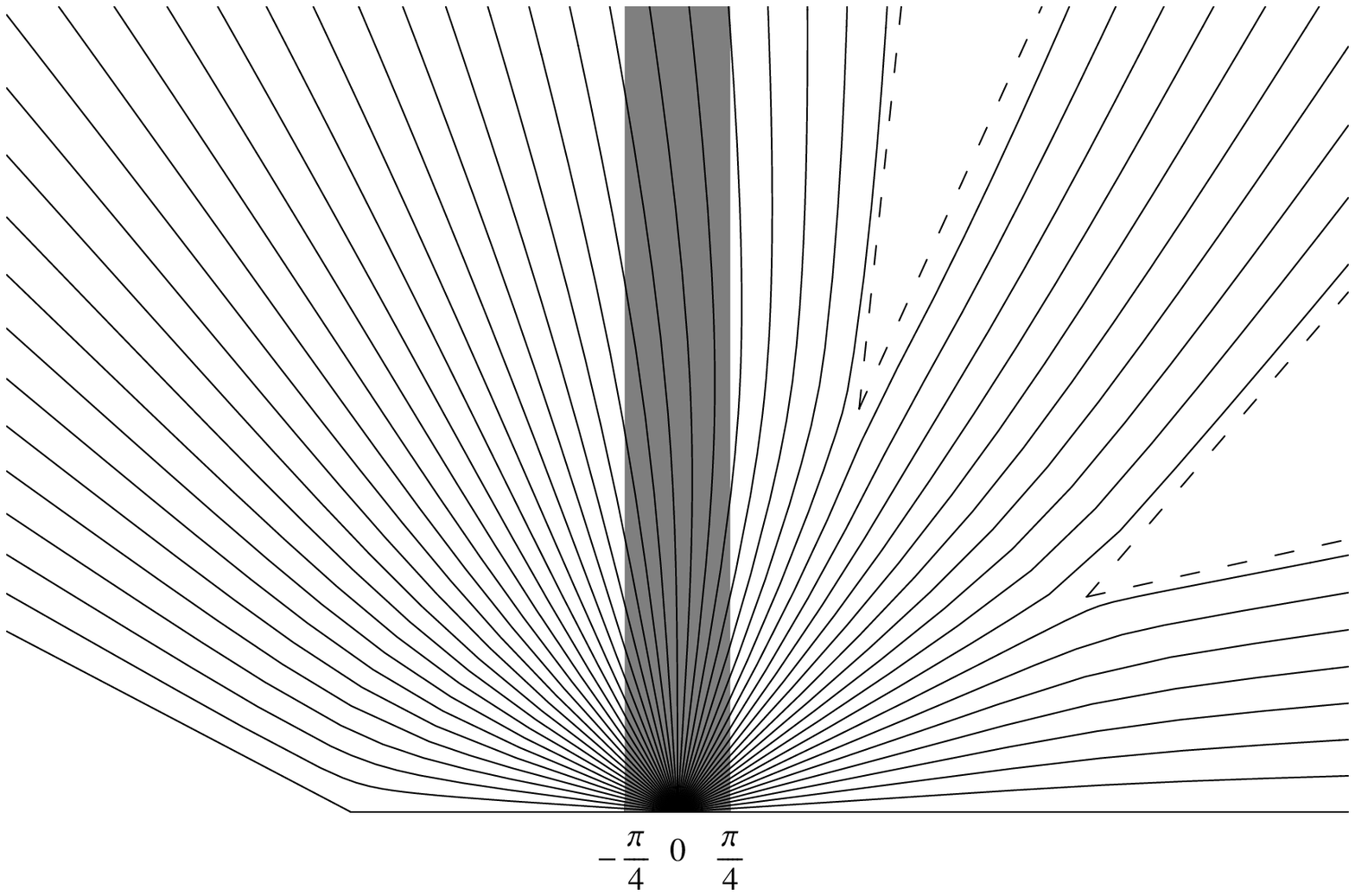,width=13cm}
\caption{The $u$ plane representation of a more ``generic''
  state. This state lives in $\HH_{39}$. The defining polynomial has
  single roots at $f=2\pm 4i$, double roots at $f=4\pm 2i$ and a triple
  root at $f=-3$. These roots produce two couples of identified
  curves, separated by angles of
  $\frac{4\pi}{39}$ and $\frac{8\pi}{39}$, and one vertex, whose
  turning angle is $\frac{6\pi}{39}$.
  The local coordinate patch is in grey.
  The lines represent the images of radial lines, separated in the $f$
  plane by an angle of $\frac{\pi}{50}$ from each other.
  Two dashed lines meeting at a point are identified.}
\label{fig:H39}
}

Star-multiplication
in $\HH_n$ is most easily performed using gluing of
half surfaces. In order to be a subalgebra we should demand the
closure of this set under star-multiplication.
It would be enough to show that given a set of vertices in the $u$
plane together with locations of
conical singularities, such that all the angle restrictions discussed
above hold for some $n$, it is possible to construct a
polynomial $P_n(f)$ of the form~(\ref{PnPSL}).

We can refer to such a map as a ``generalized Schwarz-Christoffel''
mapping. In fact, this kind of mappings were considered in the
literature, although in a different context
in~\cite{Aurell:1994ib}.
It was noted there that these functions map to polygons with conical
singularities.
However, to the best of our knowledge, a proof
of the completeness of generalized Schwarz-Christoffel maps was
never given before.
We have to prove two things,
\begin{itemize}
\item Given a convex polygon with prescribed conical singularities,
  there exists a conformal map of a properly punctured upper half
  plane onto the polygon\footnote{This is not apriori obvious,
    since the singularities prevent us from using Riemann's
    theorem.}.
\item All such conformal maps are of the generalized
  Schwarz-Christoffel type.
\end{itemize}
The number of free parameters in this problem is correct, as in the
usual Schwarz-Christoffel
case, but this fact by itself does not constitute a proof.
Finding an analytical relation between the (cut) polygon and the
polynomial coefficients
would constitute a proof, but such a relation is not know even
for the usual case.
We give a detailed proof of these assertions,
but since the proof is purely mathematical we defer it
to the appendix.

It should be stressed that the map we have is indeed to a surface with
a conical singularity, and not to surfaces before gluing, as in
figures~\ref{fig:H8_2Cuts}, \ref{fig:H39}.
Given a conical singularity on a surface we describe it using surfaces
with identifications. But there are many ways to draw the
identified curves. A change in the form of the identified curves in
the $u$ plane,
which follows by a compatible change of the cut in the $f$ plane, does
not change the conformal map. However, it is possible to change the
identified curves in the $u$ plane without modifying the cut on the
$f$ plane. The new
surface on the $u$ plane before identification and the upper half $f$
plane minus the cut,
both have disk topology, and as such are conformally equivalent. 
But these maps in general will not map the two sides of the $f$ plane
cut to equidistant points on the two $u$ plane identified curves. As
such their gluing will introduce distortions in addition to the
conical singularity. Given a cut there should exist a
unique way to ``open the conical singularity'' in the $u$ plane and
vice versa.
The generalized Schwarz-Christoffel map does not introduce
such distortions.

From a geometric point of view, the map is most properly described by
avoiding the cuts altogether, and considering
a map from the punctured upper half plane to a punctured
surface. We use this geometric description in the appendix.

\section{Conclusions}

In this paper we dealt with several issues regarding surface states
and star-subalgebras. First, we showed the equivalence of two surface
state criteria and found some analogous ghost sector criteria.
Then, we described the surface states that have
a simple representation in the continuous basis.
We showed that the twist invariant states are wedge states and
purebred butterflies and that the non-twist invariant states are
hybrid butterflies. We also elaborated on the properties of the hybrid
butterflies.
We found that all these states are described by
a simple geometric picture and discussed generalizations,
which enabled us to find many other subalgebras.
These subalgebras are built using the generalized Schwarz-Christoffel
map and contain states with conical singularities. A proof of the
generalized Schwarz-Christoffel theorem can be found in the appendix.

We could define many more subalgebras using the states we
described. For example, any set of rank one projectors, which have a
common left or right part, forms a subalgebra. As most of our states
are projectors, we could have defined many such subalgebras. As
another example we note that~(\ref{n|m}) implies that both
$\cup_{n|m} \HH_m$ and $\cup_{m|n} \HH_m$ for some given $n$ are
subalgebras. There are many other examples. What would be
interesting, though, is to find subalgebras that are big
enough for addressing a particular problem on the one hand and that
have a simple realization of the star-product on the other hand.
For example, the star-product is realized in a very simple way on
$\HH_4$, where the states are trivially manipulated both in the CFT
and in the operator language. These states are probably enough for
describing VSFT multi-D-brane solutions~\cite{Fuchs:2002zz}.
These facts are what makes $\HH_4$ an interesting object.
However, for the higher $\HH_n$ neither do we know of a simple
algebraic description of the multiplication rule (due to the
complexity of the Schwarz-Christoffel parametric problem and the lack
of a simple $\ka$-representation), nor do we know for which
problem it can be of help. It is not even clear if
surface states, or simple generalizations thereof such as surface
states with some ghost
insertions~\cite{Rastelli:2000iu,Schnabl:2002gg,Okawa:2003zc,Yang:2004xz},
are large enough for the problem of finding analytic
string field theory solutions.

Conical singularities contribute delta function curvature
singularities~\cite{Witten:1986cc,Banados:1992wn}.
The bosonized ghost action implies that there are ghost insertions at
such points, as noted for string field theory vertices
in~\cite{Witten:1986cc}.
It may be possible to simplify the star-product of states with conical
singularities by using these facts.
Finding a simple
representation of the star-product for our spaces would promote their
practical use. Another possible direction would be to characterize the
subalgebras in terms of the tau-function using the tools
of~\cite{Boyarsky:2002yh,Bonora:2002un,Boyarsky:2003kr}.
It is also possible to form star-subalgebras by using the Virasoro
operators $K_n$~\cite{Gaiotto:2002kf}. It would be interesting to
check if our spaces can be related to these ones, or to other
star-derivations.

\section*{Acknowledgments}

We would like to thank Ofer Aharony, Gidi Amir, Paul Biran,
Ori Gurel-Gurevich, J\"{u}rg K\"{a}ppeli, Alon Marcus, Yaron Oz, Misha
Sodin, Jacob Sonnenschein, Stefan Theisen and Barton Zwiebach
for discussions.
M.~K. would like to thank the Albert Einstein
Institute, where part of this work was performed
for hospitality.
This work was supported in part by the German-Israeli Foundation for
Scientific Research.

\appendix
\section{Generalizing the Schwarz-Christoffel proof}

Standard proofs of the Schwarz-Christoffel formula, as the ones
in~\cite{Ahlfors,Aurell:1994ib} are hard to generalize for the case with
conical singularities, since they rely heavily on specific properties
of meromorphic functions.
We therefore, provide an alternative proof
in~\ref{sec:ASC}, which uses only
topological arguments to prove that the naive parameter counting is
indeed adequate. We examine only our case of interest, namely the case
with one vertex at infinity, and so, when we refer to a polygon we
actually mean a polygon with one vertex at infinity.
However, we do not limit the discussion to polygons with a well
defined local coordinate patch.
Then, in~\ref{sec:AGSC}, the proof is
extended to the case of a
generalized Schwarz-Christoffel map with one vertex at infinity.

\subsection{The Schwarz-Christoffel map}
\label{sec:ASC}

Given the angles of the polygon, we know the exponents of the
prevertices.
All Schwarz-Christoffel maps with the correct number of prevertices
that have adequate exponents give polygons with the correct
angles. What we want to prove is that all polygons with these angles
may be described by such maps. The proof is by induction on the
number of vertices.
We fix the $\PSL(2,\RR)$ freedom by looking for maps for
which~(\ref{uoff}) reduces to
\begin{equation}
\label{uSCproof}
u=e^{i\theta} \int_0^f d f \prod_{i=1}^m
\left(\frac{1+|f_i|}{f_i- f}\right)^{\!\beta_i}\,.
\end{equation}
Here we set $u_0=0$, which amounts to
\begin{equation}
\label{f0SCproof}
f(0)=0\,.
\end{equation}
We also set
$f(\infty)=\infty$ by demanding
\begin{equation}
\label{betaLimit}
\sum_{i=1}^m \beta_i < 1\,.
\end{equation}
While in the general case this inequality can be saturated, all
polygons that saturate this inequality can be regarded as limits of
polygons which do
not\footnote{\label{foot2}With a definition of the limit given shortly,
the singular polygon with a single $\beta=1$ vertex (in addition to the
$\beta=1$ vertex at infinity) is an exception. However, it is clear
how to represent it as a Schwarz-Christoffel map, it is
the $\tanh(c u)$ solution, which we discarded in
section~\ref{sec:SSinHka}. It is not a polygon in a strict sense in
any case, since the ``finite vertex'' is also at infinity.}. 
This limit procedure is well defined in both the $f$ and $u$ planes.
Thus, there is no loss of generality in the
assumption~(\ref{betaLimit}). The scale constant $c$ of~(\ref{uoff})
was set in~(\ref{uSCproof}) in a particular way that leaves out only a
phase, which takes care of the polygon orientation.
It is a matter of straightforward algebra to see that it is always
possible to get to such a form with a $\PSL(2,\RR)$ rescaling. 
In our case of interest we have
$\beta_i=\frac{2 k_i}{n}$ (all powers are rational and the angles are
rational multiples of $\pi$). In fact, for our case $\theta$ is also
known, since our assumption~(\ref{f0SCproof}) also implies that the
image of a neighborhood of the $f$ plane origin should be mapped to
an open interval around the $u$ plane origin in an orientation
preserving way, that is $f'(0)>0$. We therefore deduce that
\begin{equation}
\theta=\pi\sum_{f_i<0} \beta_i\,.
\end{equation}
Our proof is not restricted to $\HH_\infty$ surfaces. The
inclusion of polygons whose boundary does not include the origin is
also straightforward.

The most natural decomposition of the space of polygons consists of
subspaces of all polygons with a given number of vertices and with
given turning angles in these vertices. The vertices in such a
subspace are naturally ordered, with the vertex at infinity as the
first one. However, we find it easier to consider
all subspaces with a given set of angles together, regardless of the
order of vertices, other than the one at infinity.
We also allow in such a space
the merging of vertices. For example, we consider a polygon with a
single angle of $\beta_1+\beta_2$ as being part of the space of
polygons with the two angles $(\beta_1,\beta_2)$. We call this space
$\U_{\vec\beta}$, with $\vec\beta=(\beta_1,\beta_2)$. Generally we have
a natural embedding $\U_{\vec\beta_a}\rightarrow \U_{\vec\beta_b}$,
provided the elements of $\vec\beta_b$ can be grouped into partitions
of elements of $\vec\beta_a$. Thus, the decomposition we consider is
not a disjoint one.

The set of Schwarz-Christoffel maps that map the upper half plane to
polygons in $\U_{\vec\beta}$ is denoted by
$\F_{\vec\beta}$. This space is topologically $\RR^m$,
where the $i^{\text{th}}$ component equals to $f_i$
in~(\ref{uSCproof}). In principle we should have divided this space by
symmetry factors of permutation groups of equal angles.
However, it is easier to pretend that these prevertices are
distinguishable and at the same time distinguish equivalent vertices
in the definition of $\U_{\vec\beta}$.
Proving the assertion for these covers of
the correct spaces is enough, as any representative for the
conformal map in $\F_{\vec\beta}$ will do the job.
Moreover, given two conformal maps from the upper half plane to a
given polygon, $u_1(f),u_2(f)$, the composition $u_2^{-1}\circ u_1$ is
a conformal map from the upper half plane to itself and so is in
$\PSL(2,\RR)$, but since we already fixed this ambiguity, this map is
actually the identity map and we have $u_1=u_2$.
The topology of $\U_{\vec\beta}$ is also $\RR^m$. Here, the
$i^{\text{th}}$ component, $u_i$, equals to the oriented distance along
the boundary of the polygon from the origin to the $i^{\text{th}}$
vertex.

There is a natural injection
\begin{equation}
\label{FUmap}
\F_{\vec\beta}\rightarrow\U_{\vec\beta}\,,
\end{equation}
where a given map $u(f)\!\in\F_{\vec\beta}$ is sent to the polygon,
which constitutes its range.
From~(\ref{uSCproof}), (\ref{betaLimit}) it is clear that this map is
continuous with the topologies defined above. These definitions of
topologies also settle the limit issues mentioned
below~(\ref{betaLimit}).
We now turn to prove that this injection is also onto.
The induction base is trivial, since with zero vertices there is only
a map from the upper half plane to itself, and it is trivially of the
Schwarz-Christoffel type.

Now, let $\vec \beta=(\beta_1,..,\beta_m)$, and let $u_i$ approach the
boundary of $\U_{\vec\beta}$ for some $1\leq i \leq m$. The approach
to the boundary is well defined, and the limit polygon belongs to
$\U_{\vec\beta_o^i}$, where we define
\begin{equation}
\vec\beta_o^i \equiv (\beta_1,..,\hat \beta_i,..,\beta_m)\,,
\end{equation}
and as usual $\hat\beta_i$ means that this factor is omitted.
By the induction hypothesis the natural map
$\F_{\vec\beta_o^i}\rightarrow\U_{\vec\beta_o^i}$ is a bijection.
Moreover, due to the form of~(\ref{uSCproof}), the space
$\F_{\vec\beta_o^i}$ is a continuous limit of $\F_{\vec\beta}$.
The above is true for all $i$.
We can, therefore, compactify the spaces $\U_{\vec\beta}$,
$\F_{\vec\beta}$, such that the compact spaces
$\overline{\U_{\vec\beta}}$, $\overline{\F_{\vec\beta}}$ are
topologically closed cubes, and extend~(\ref{FUmap})
continuously to the closure,
\begin{equation}
\overline{\F_{\vec\beta}}\rightarrow
 \overline{\U_{\vec\beta}}\,.
\end{equation}
This is true also for lower dimensional boundary components in a
cellular decomposition of our spaces,
$\overline{\U_{\vec\beta}}$, $\overline{\F_{\vec\beta}}$.
These can be realized as multiple limits of vertices
or prevertices.

We showed that by the induction hypothesis, every lower dimensional
cell in the boundary of $\overline{\F_{\vec\beta}}$ is mapped
bijectively onto the corresponding boundary
of $\overline{\U_{\vec\beta}}$. 
Thus, we showed that when restricted to the boundary, our map is
homotopic to the identity map.
We can now use the fact that our space is contractable in order to
invoke standard topological arguments and prove that the map
is onto.
Our spaces are homeomorphic to manifolds, so there is no loss of
generality by using arguments about manifolds.
It is known that a map $\partial X\rightarrow Y$, with $X,Y$
manifolds, that can be extended to a map $X\rightarrow Y$ is
of zero rank. Now, suppose that the map
$\overline{\F_{\vec\beta}}\rightarrow \overline{\U_{\vec\beta}}$ is
not onto, and let $u_0$ be a point not in the range. Since
$\overline{\U_{\vec\beta}}$ can be contracted to $u_0$, we can define
a map $\overline{\F_{\vec\beta}}\rightarrow \partial \U_{\vec\beta}$
with the same boundary value as before. This shows that the boundary
map is of zero rank, which contradicts its homotopy to the identity.
It therefore follows, that the map is
a bijection and the Schwarz-Christoffel theorem follows.

\subsection{Generalized Schwarz-Christoffel map}
\label{sec:AGSC}

We now turn to generalize this proof for the case of a generalized
Schwarz-Christoffel map. The relevant spaces in this case are
$\U_{\vec\beta,\vec\gamma}$, $\F_{\vec\beta,\vec\gamma}$, with
$\vec\gamma=(\gamma_1,..,\gamma_n)$ describing the angle deficits of
$n$ conical singularities on the polygon.
We fix the $\PSL(2,\RR)$ symmetry in a manner similar to the one used
in~(\ref{uSCproof}), by writing
\begin{equation}
\label{uGSCproof}
u=e^{i\theta} \int_0^f d f \prod_{i=1}^m
  \left(\frac{1+|f_i|}{f_i-f}\right)^{\!\beta_i}
 \prod_{j=1}^n
  \left(\! \frac{\big(1+|\tilde f_j|\big)^{\gamma_j}}
    {\big((\tilde f_j-f)(\tilde f_j^*-f)\big)^{\gamma_j/2}}
 \! \right).
\end{equation}
Here $f_i$, for $1\leq i \leq m$ are the prevertices, which are bound
to the real axis. The $\tilde f_j$, with $1\leq j \leq n$ are the
pre-singularities, which take values in the upper half plane, and
$\tilde f_j^*$ are their complex conjugates.
Again $f(0)=0$, $f(\infty)=\infty$ and the rescaling factor in
uniquely fixed.

We adopt the geometric
picture of not presenting the $f$ plane cuts, and $u$ plane
identified curves explicitly.
Geometrically there is no dependence on the form and direction of the
identified curves, as long
as they do not introduce distortions other than deficit angles.
They can go one through another, and pass
over conical singularities. They can end at infinity, or on the finite
boundary. To illustrate this point we show in fig.~\ref{fig:cuts} a
simple case of a polygon with a single vertex and a single conical
singularity, with three different representations of the identified
curves.
\FIGURE{
\epsfig{figure=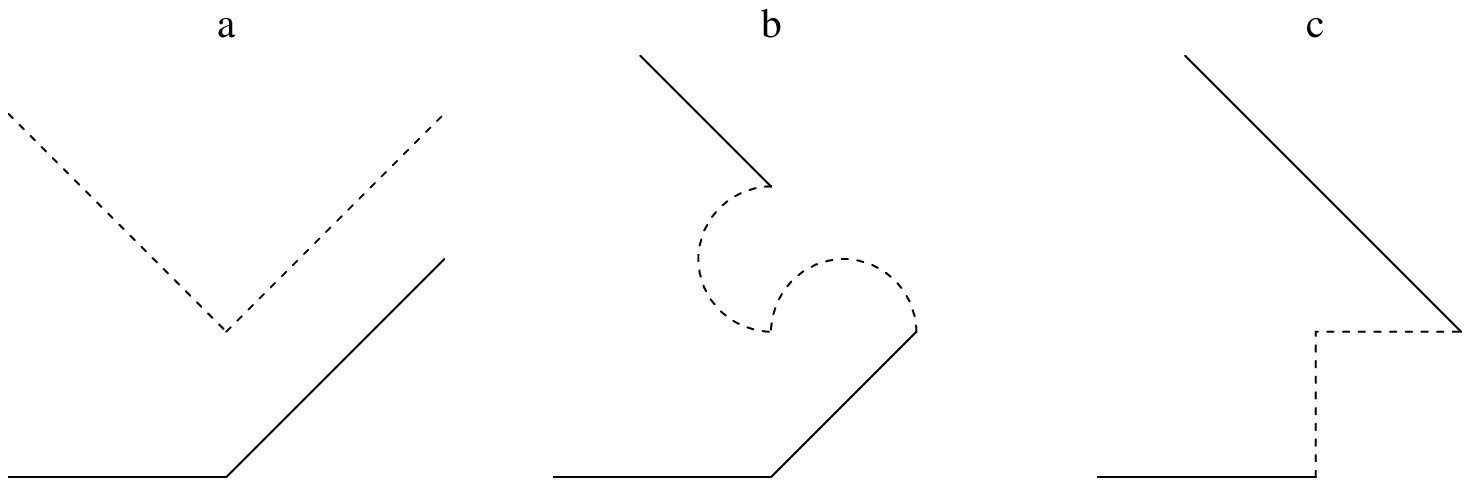,width=15cm}
\caption{A polygon with a single vertex of $\beta=\frac{1}{4}$, and a
single conical singularity of $\gamma=\frac{1}{2}$, represented in
three equivalent ways that differ by the locations of the
(dashed) identified curves.
In $a$, the identified curves are two lines going from the conical
singularity to infinity. In $b$, the identified curves consist of two
semi-circles, which go to a point on the boundary. In $c$, the
identified curves are again two lines, which now go to the
vertex. There are many other possibilities for describing this
surface.}
\label{fig:cuts}
}

The condition that the total deficit angle is less than $\pi$, which
is the same as the condition $f(\infty)=\infty$,
generalize~(\ref{betaLimit}) to,
\begin{equation}
\label{BetaGammaLimit}
\sum_{i=1}^m \beta_i + \sum_{j=1}^n \gamma_j< 1\,.
\end{equation}
Again, this condition should in principle be
saturated, but as mentioned above, there is no loss
of generality by keeping it as is. The only new polygons that we miss
now in addition to what was mentioned in footnote~\ref{foot2} are the
wedge states, and we already know their description in terms of
generalized Schwarz-Christoffel maps.

As before, we distinguish equivalent points in the spaces
$\U_{\vec\beta,\vec\gamma}$, $\F_{\vec\beta,\vec\gamma}$.
It is clear, that the space $\F_{\vec\beta,\vec\gamma}$ has the
topology $\RR^m\otimes H^n$, where $H$ is the open upper half plane.
For the topology of $\U_{\vec\beta,\vec\gamma}$, we consider a fixed
value for $\vec\beta$ and add the conical singularities one at a
time. As long as~(\ref{BetaGammaLimit}) is maintained, we have an open
topological disk, that is $H$, for the position of the next conical
singularity. The space over a given $\vec\beta$ is, therefore,
$H^n$. Thus, $\U_{\vec\beta,\vec\gamma}$ carries a topology of a
fibration of $H^n$ over $\RR^m$, but since the last space is
homotopically trivial, the topology is $\RR^m\otimes H^n$, as it is
for $\F_{\vec\beta,\vec\gamma}$.

We now perform a double induction in $(m,n)$. The pairs are
$\omega^2$-well-ordered according to,
\begin{equation}
\label{inductionOrder}
(m_1,n_1)\prec(m_2,n_2) \Longleftrightarrow
   (n_1 < n_2) \text{ or } (n_1=n_2 \text{ and } m_1<m_2)\,.
\end{equation}
The induction base is the same as before.
Let $(m,n)$ and an adequate $\vec\beta,\vec\gamma$ be given.
The spaces $\U_{\vec\beta,\vec\gamma}$, $\F_{\vec\beta,\vec\gamma}$
are naturally compactified as before. For the vertices and
prevertices the boundaries are at $\pm \infty$ as before.
We want to examine the boundary of the positions of conical
singularities. The boundary of $H$ is the circle
that consists of the real line and the point at infinity. For a
polygon, let a conical singularity approach the point at
infinity. Then we see that the limit is a polygon with one less
singularity. The same conclusion can be drawn in
$\F_{\vec\beta,\vec\gamma}$ due to~(\ref{uGSCproof}).
Now, let the point approach the real line. For the polygon it is seen
from fig.~\ref{fig:cuts} that a new vertex is created with
$\beta_{i+1}=\gamma_j$. Thus, the boundary of
$\U_{\vec\beta,\vec\gamma}$ is covered with the spaces
$\U_{\vec\beta^{(i,j)},\vec\gamma_o^j}$, with
\begin{align}
\vec\gamma_o^j&=(\gamma_1,..,\hat \gamma_j,..,\gamma_n)\,,\\
\vec\beta^{(i,j)}&=(\beta_1,..,\beta_m,\gamma_j)\,.
\end{align}
The space $\RR^m\otimes H^n$ is homeomorphic to $\RR^{m+2n}$ in a way
that is consistent with our definition of boundaries. By the induction
hypothesis we get a non-trivial map of the boundaries, and the
generalized Schwarz-Christoffel theorem follows.

Our proofs can be further generalized in several ways.
One possible generalization is to the case without a vertex at
infinity. 
Another generalization is to the case with arbitrary complex
roots. In this case there should be no distinction between the real
and non-real roots. Both should be considered as sources of conical
singularities, and the map should be considered as a map of Riemann's
sphere minus some points to a topological sphere with a complex
structure that has some prescribed conical singularities.

\bibliography{FK}

\end{document}